\documentclass[nofootinbib,preprint,amsmath,amssymb,showpacs,showkeys,aps,longbibliography]{revtex4-2}
\usepackage{graphicx}
\usepackage{placeins}
\usepackage[utf8,latin1]{inputenc}
\usepackage{dcolumn}
\usepackage{mathrsfs}
\usepackage{subcaption}
\usepackage{gensymb}
\usepackage{tensor}
\usepackage{academicons}
\usepackage{mathtools, nccmath}
\usepackage[overload]{empheq}
\usepackage{tikz,xcolor}
\usepackage{cases}
\usepackage{setspace}
\usepackage{lipsum}
\usepackage[normalem]{ulem}
\usepackage{bm}
\usepackage{soul}
\usepackage{hhline}
\usepackage{bigints}
\usepackage{amsmath}
\usepackage[T1]{fontenc}
\usepackage{fancyhdr}
\usepackage{hyperref}
\usepackage{combelow}
\usepackage{lmodern}
\usepackage{hyperref}
\usepackage{latexsym,amsmath,amscd,amssymb,graphics,mathrsfs}
\usepackage{graphicx}
\usepackage{color}
\usepackage{ulem}
\usepackage[all]{xy}
\usepackage{framed}
\usepackage{float}
\usepackage{mathtools} 
\usepackage{url}
\usepackage{xcolor}
\usepackage{wrapfig}
\usepackage{setspace}
\usepackage{tikz}
\usepackage{enumitem}
\usepackage{ragged2e}
\usepackage{physics}
\usepackage{tikz-3dplot}
\usetikzlibrary{decorations.pathmorphing}
\usetikzlibrary{patterns}
\hypersetup{colorlinks, linkcolor={red},citecolor={blue},urlcolor={blue}}  

\begin{document}
	
	\title{Complex-time parametrization of evanescent tunneling and transmission delay in non-Hermitian scattering}
	
	\author{Daia Iulia-Maria$^{1}$} 
	\email{iulia.daia03@e-uvt.ro}
	
	\affiliation{$^1$ Department of Physics, West University of Timi\cb{s}oara \\ Bd. Vasile P\^arvan 4, Timi\cb{s}oara 300223, Romania	}
	
	\date{\today} 
	
	\begin{abstract}
		
		The quantum tunneling process is a fundamental concept that presents the unconventional nature of quantum dynamics. The main anomaly of this effect revolves around the choice of imposing the classical 4-momentum to become imaginary during the barrier traversal. This paper addresses this paradox by introducing a complex-time parametrization, which uses a complex temporal coordinate, $\tau = t + i\kappa$. Here, the real axis $(t)$ represents standard thermodynamic time, while the imaginary axis $(i\kappa)$ is correlated to spatial evanescence. We demonstrate that evaluating this extended complex-time framework yields a hyperbolic section of the causal lightcone, proving that the traversal is strictly bounded by the speed of light and preserves causal disconnection. Applied to quantum tunneling, we resolve the anomaly of imaginary momentum through a continuous geometric signature flip of the effective complex-time momentum norm from a timelike to a spacelike state. This theoretical model finds a physically phenomenological analogy in microwave scattering experiments using open, non-Hermitian systems, which link complex transmission time delay to spatial wave decay.
		
		\vspace{0.5cm}
		
	\end{abstract}
	
	\maketitle
	
	\newpage
	
	\section{Introduction}\label{Sec:Introd}
	
	Quantum tunneling represents one of the most fundamental phenomena studied in quantum mechanics. It is one of the hallmarks of this field and describes the unconventional behavior of non-classical particles. This phenomenon enables particles to traverse barriers which are considered impenetrable by classical physics. Despite its theoretical discovery in 1927 by Friedrich Hund \cite{Hund1927}, there is still an ongoing debate around it, regarding the type of time spent, its duration and experimental measure \cite{Steinberg:1994ks, Camus:2016yhc, Popov:2005rp}, and the specific mechanics inside the barrier \cite{Pisanty:2014qre, Klaiber:2022ogb}.
	
	Another open question is about quantum jumps. While early theories assumed the transition inside the barrier is instantaneous and bypasses the speed of light $c$ limit, modern experiments \cite{Schach:2024kvq, Demir:2017qkj} show that the particle moves gradually from one state to another; therefore, the leap unfolds continuously rather than infinitely fast.
	
	In this paper, we present an alternative approach which introduces the complexification of time in order to solve the anomalies of superluminal traversal and imaginary 4-momentum governing the tunneling process. We propose extending linear time to a complex plane: $\tau = t + i \kappa$. This phenomenological parametrization relies on three mathematical particularities: the co-existence of real and imaginary temporal states, the non-perpendicularity of their dynamic coupling, and the strict degeneracy of the resulting complex-time parametrization metric. We demonstrate that transforming the classical $(3+1)$ lightcone into a lighthyperbola proves that the mathematical traversal along the imaginary axis is strictly bounded by the universal speed limit $c$, ensuring causality is preserved. In addition, we prove that barrier penetration is not a discontinuous quantum jump, but involves a continuous rotation of the temporal axis governed by a geometric phase transition angle $\theta$, without violating the speed of light.
	
	This paper is structured as follows. We introduce the general formula of the complex-time parametrization. Then we derive the complex-time metric, its associated extended vector formalism, and the transformation of the lightcone into the lighthyperbola. We apply this framework to geometrically show that barrier penetration is governed by an effective complex-time momentum norm signature flip that eliminates the imaginary 4-momentum paradox. Ultimately, we present empirical data from non-Hermitian microwave experiments that provides a phenomenological analogy for an imaginary component of a response time.
	
	\section{Complex-Time Parametrization}\label{Sec:twotime}
	
	We define the complex-time parameter:
	
	\begin{equation}
		\tau = t + i \kappa
		\label{tau-eq}
	\end{equation}
	
	The variable $t$ represents the real time. We will refer to this as "thermodynamical time", due to the fact that it can be measured by a standard clock and it is governed by the Second Law of Thermodynamics. Real thermodynamic time is strictly sequential, evolving unidirectionally through past, present, and future states.
	
	The imaginary element $i$ acts as a geometric rotating operator indicating orthogonality, while the term $\kappa$ is an auxiliary parameter. As we will demonstrate in Sec. \ref{Sec:tunneling}, this imaginary-time coordinate serves as a bookkeeping representation of spatial evanescence in quantum tunneling. Furthermore, the term $i\kappa$ introduces an orthogonal temporal axis. The nature of this term is fundamentally different from thermodynamical time $t$, as it does not obey the laws of thermodynamics and possesses no arrow of time. These fundamental properties will be proved in the following sections.
	
	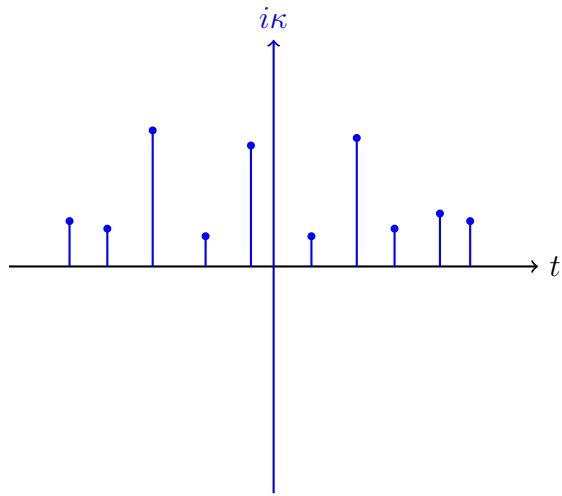
\begin{figure}[htbp]
		\centering
		\begin{tikzpicture}
			
			\draw[->, thick] (-3.5, 0) -- (3.5, 0) node[right] {$t$};
			
			\draw[->, thick, blue!80!black] (0, -3) -- (0, 3) node[above, text=blue!80!black] {$i\kappa$};
			
			\draw[blue, thick] (-2.7, 0) -- (-2.7, 0.6); \fill[blue] (-2.7, 0.6) circle (1.5pt);
			\draw[blue, thick] (-2.2, 0) -- (-2.2, 0.5); \fill[blue] (-2.2, 0.5) circle (1.5pt);
			\draw[blue, thick] (-1.6, 0) -- (-1.6, 1.8); \fill[blue] (-1.6, 1.8) circle (1.5pt);
			\draw[blue, thick] (-0.9, 0) -- (-0.9, 0.4); \fill[blue] (-0.9, 0.4) circle (1.5pt);
			\draw[blue, thick] (-0.3, 0) -- (-0.3, 1.6); \fill[blue] (-0.3, 1.6) circle (1.5pt);
			
			\draw[blue, thick] (0.5, 0)  -- (0.5, 0.4);  \fill[blue] (0.5, 0.4)  circle (1.5pt);
			\draw[blue, thick] (1.1, 0)  -- (1.1, 1.7);  \fill[blue] (1.1, 1.7)  circle (1.5pt);
			\draw[blue, thick] (1.6, 0)  -- (1.6, 0.5);  \fill[blue] (1.6, 0.5)  circle (1.5pt);
			\draw[blue, thick] (2.2, 0)  -- (2.2, 0.7);  \fill[blue] (2.2, 0.7)  circle (1.5pt);
			\draw[blue, thick] (2.6, 0)  -- (2.6, 0.6);  \fill[blue] (2.6, 0.6)  circle (1.5pt);			
			
		\end{tikzpicture}
		\caption{Visualization of time complexification: the x-axis $(t)$ represents the linear real time; the y-axis $(i\kappa)$ represents the auxiliary parameter}
		\label{fig:complex_time}
	\end{figure}
	
	It is essential to distinguish the complex-time coordinate expansion from the concept of imaginary time used in Quantum Field Theory (QFT). In QFT, imaginary time is introduced via a Wick rotation \cite{Peskin:1995ev}, which is a mathematical tool that turns real time into imaginary time $(t \rightarrow -i\tau)$. Instead of a time coordinate substitution, this framework performs a time expansion, promoting time to a complex plane. In this geometry, the co-existence of thermodynamical time $t$ and the auxiliary term $i\kappa$ represents a fundamental particularity of this framework; rather than replacing real standard time, $\kappa$ operates to parametrize spatial evanescence.
	
	Due to the complexification of time, the general formula shown in Eq. \eqref{tau-eq} can be written as:
	 
	 \begin{equation}
	 	\tau = t + i\kappa = \text{Re}(\tau) + i\text{Im}(\tau)
	 	\label{real_imaginary_part}
	 \end{equation}
	
	Taking into account the Principle of Dimensional Homogeneity, all terms must be measured in identical units. The standard SI unit of measurement for thermodynamical time $t$ is seconds, which dictates that the orthogonal auxiliary parameter $\kappa$ must also share this physical dimension. As we will prove in Sec. \ref{Sec:invariant} (through the line element derivation) and \ref{Sec:tunneling} (through the $\kappa(x)$ spatial decay mapping), $\kappa$ is a rigorous measure of temporal duration (seconds).
	
	However, from a phenomenological perspective, $\kappa$ can not be measured using standard chronological instruments like real thermodynamical time $t$. Its value is instead derived from physical spatial observables: the spatial penetration distance $(x)$, the decay constant $(\chi)$ and the angular frequency $(\omega)$ of the particle during quantum traversal.
	
	Since time has been promoted from a linear dimension to a complex plane, we apply Euler's relation to express the complex-time parameter in polar form:
	
	\begin{equation}
		\tau = \left| \tau \right| \cdot e^{i\theta} = \left| \tau \right| \cdot [cos (\theta) + i sin (\theta)]
		\label{polar_form}
	\end{equation}
	
	Physically, $\left| \tau \right|$ is the invariant complex-time magnitude. The thermodynamical time $t$ and the term $i\kappa$ are geometric projections of this absolute magnitude, dictated by the phase angle $\theta$. We introduce the phase rotation angle $\theta$ as a transition parameter that mathematically dictates the shift from the classical region - a pure oscillatory state where the kinetic energy is positive - to the quantum region, representing the evanescent state where the kinetic energy becomes negative. It is crucial to observe that we do not treat $t$ and $\kappa$ as separate, independent coordinates; they are linked through this phase angle.
	
	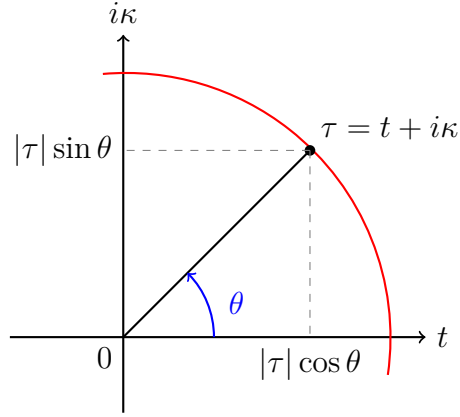
\begin{figure}[htbp]
		\centering
		\begin{tikzpicture}
			\draw[->, thick] (-1.5, 0) -- (4, 0) node[right] {$t$};
			\draw[->, thick] (0, -1) -- (0, 4) node[above] {$i\kappa$};
			\node[below left] at (0,0) {$0$};
			
			\draw[thick, red] (3.5, -0.5) arc (-8.13:95:3.5);
			
			\coordinate (P) at (2.47, 2.47); 
			\draw[thick, black] (0,0) -- (P);
			\fill[black] (P) circle (2pt) node[above right] {$\tau = t + i\kappa$};
			
			\draw[dashed, gray] (P) -- (2.47, 0) node[below, text=black] {$|\tau|\cos\theta$};
			\draw[dashed, gray] (P) -- (0, 2.47) node[left, text=black] {$|\tau|\sin\theta$};
			
			\draw[thick, blue, ->] (1.2, 0) arc (0:45:1.2);
			\node[blue, right] at (1.25, 0.45) {$\theta$};
		\end{tikzpicture}
		\caption{Geometric representation of the complex-time parametrization. The macroscopic thermodynamical time $t$ and the auxiliary parameter $i\kappa$ are connected through the phase transition angle $\theta$. This parameter describes the continuous shift from a pure oscillatory state (real axis) to an evanescent state (imaginary axis). }
		\label{fig:bitemporal_rotation}
	\end{figure}

	The phase angle must be restricted to the fourth quadrant of the trigonometric circle $(0 \geq \theta \geq -\frac{\pi}{2})$ to ensure thermodynamical time moves forward $(t \geq 0)$. Moreover, during the quantum tunneling process, the particle's wavefunction progressively decays. To map this geometrically, the framework restricts the orthogonal term to traverse along the negative axis $(\kappa \leq 0)$.
	
	Ultimately, it is essential to clarify why this framework limits its complexification exclusively to time. Extending all spacetime coordinates into a complex or pseudo-complex domain as shown in Ref. \cite{Hess:2008wd} would add unnecessary degrees of freedom without physical justification for this phenomenological model. More importantly, real thermodynamical time is intrinsically asymmetric, unidirectional and irreversible, whereas standard $3D$ space is isotropic and bidirectional. Because real space already permits symmetric translation without violating thermodynamic laws, there is  neither physical nor mathematical justification for expanding real space into an imaginary domain.
	
	\section{Standard (3+1) Spacetime}\label{Sec:normalspacetime}
	
	In this section, we present the fundamental basics of standard $(3+1)$ spacetime, which serves as our benchmark geometry, in the sense that it allows us to understand better the complexified concepts we will later encounter in the paper.
	
	Our starting point is the metric tensor $g_{\mu\nu}$, before passing to the extended complex-time parametrization framework. 
	
	\subsection{4D Metric Tensor} \label{Sec:4d_metric}
	
	In standard $4D$ Special Relativity, spacetime geometry is defined by the metric tensor $g_{\mu\nu}$, which is used to calculate distances and time intervals in flat spacetime \cite{Carroll:1997ar}. Greek indices span over four dimensions and range as $\alpha,\beta,\dots =0,\dots,3$. 
	
	In real Minkowski spacetime, the covariant metric tensor $g_{\mu\nu}$ is a $4 \cross 4$ diagonal matrix. Using the coordinates $x^0 = ct, x^1 = x, x^2 = y, x^3 = z$, the matrix is written as:
	
	\begin{equation}
		g_{\mu\nu} = \begin{pmatrix} -1 & 0 & 0 & 0 \\ 0 & 1 & 0 & 0 \\ 0 & 0 & 1 & 0 \\ 0 & 0 & 0 & 1 \end{pmatrix}
	\end{equation}
	
	The "mostly plus" metric signature is $(-,+,+,+)$. The positive sings correspond to spatial dimension $(x,y,z)$. The negative sign allocated to the $g_{00}$ component is of absolute importance, since it creates the chronological arrow of time, enforcing a cause-and-effect sequence.
	
	By definition, the contravariant form $g^{\mu\nu}$ is the inverse of the covariant one. Given that the covariant one is a pure-diagonal matrix with $-1$ and $+1$ elements, its inverse is identical to the original matrix. Therefore, the upper index form is:
	
	\begin{equation}
		g^{\mu\nu} = \begin{pmatrix} -1 & 0 & 0 & 0 \\ 0 & 1 & 0 & 0 \\ 0 & 0 & 1 & 0 \\ 0 & 0 & 0 & 1 \end{pmatrix}
	\end{equation}
	
	In standard $4D$ spacetime, the transition between the covariant and contravariant forms is straightforward, because the Minkowski metric is non-degenerate. This invertibility is essential for raising and lowering indices when manipulating 4-vectors and tensors. In contrast to this standard metric, the extended $5 \cross 5$ metric tensor using the complex-time parametrization yields a strictly degenerate metric, which prevents standard tensor inversion, requiring us to adapt the algebraic rules of the framework. 
	
	\subsection{MTW Imaginary Time Expansion}\label{Sec:MTW}
	
	In the past, physicists attempted to simplify the geometry of spacetime by introducing an imaginary time coordinate, defined as $x^4 = ict$.
	
	However, Misner, Thorne and Wheeler \cite{Misner:1973prb} famously rejected this convention for several geometric reasons.
	
	Replacing real time with an imaginary complex coordinate conceals the fundamental Lorentzian-Minkowski $(-,+,+,+)$ metric signature and forces the geometry to resemble a Euclidian $(+,+,+,+)$ space, where the zero distance between two points are exactly the same point. Moreover, this complex coordinate spares the distinction between the contravariant vectors and covariant 1-forms.
	
	The extended complex-time parametrization avoids these issues. The complexification of time implies that the $x^0 = ct$ coordinate remains intact and the real part of the metric strictly preserves the $(-,+,+,+)$ signature. The auxiliary coordinate $x^4 = c \kappa$ is assigned to parametrize quantum evanescence.
	
	Consequently, the addition of this orthogonal auxiliary variable will still preserve the fundamental causal boundaries and maintains the distinction between quantities with upper and lower indices demanded by standard relativity. 
	
	\section{Extended Complex-Time Parametrization}\label{Sec:parametrization}
	
	This section provides the mathematical and conceptual clarifications of the extended complex-time parametrization framework. Besides deriving the extended metric tensor $g_{\mu\nu}$ and the invariant interval $ds^2$, we introduce a 5-Component Parametric Formalism, which will be applied to explain the quantum tunneling regions in Sec. \ref{Sec:tunneling} .
	
	By expanding linear time into a complex plane, the framework retains the exact algebraic structure of $4D$ Minkowski spacetime. Hence, there is no need to alter the fundamental equations regarding relativistic kinematics or the definitions of spacetime causal regions.
	
	\subsection{The Extended Metric Tensor}\label{Sec:extendedmetric}
	
	At first glance, the complexification of time seen in Eq. \eqref{tau-eq} might imply that the $4 \cross 4$ diagonal matrix changes into a purely diagonal $5 \cross 5$ matrix.
	
	If we assume there is no connection between real thermodynamical time $t$ and auxiliary parameter $\kappa$ and consider the coordinates: $x^0 = ct, x^1 = x, x^2 = y, x^3 = z, x^4 = c\kappa$, the uncoupled metric tensor will be: 
	
	\begin{equation}
		g_{\mu\nu}^{\text{uncoupled}} = \begin{pmatrix} -1 & 0 & 0 & 0 & 0 \\ 0 & 1 & 0 & 0 & 0\\ 0 & 0 & 1 & 0 & 0 \\ 0 & 0 & 0 & 1 & 0  \\ 0 & 0 & 0 & 0 & 1 \end{pmatrix}
		\label{uncoupled}
	\end{equation}
	
	However, this pure diagonal matrix presents a physical problem and contradicts Eq. \eqref{polar_form}: it isolates real time $t$ from the auxiliary parameter $\kappa$.
	
	As we established before, one of the properties of this extended framework is the co-existence of both parameters, which are linked through the phase rotation angle $\theta$. Furthermore, empirical evidence which will be detailed in Sec. \ref{Sec:empirical} demands these two variables interact with one another.
	
	Consequently, this mathematical interaction requires the extended metric tensor to have off-diagonal terms. We derive these necessary imaginary cross-terms from the geometry rather than adding them arbitrary. 
	
	The focus of the next section is to calculate the complex Minkowski line element which will generate the correct coupled metric tensor.
	
	\subsection{The Complexified Invariant Interval}\label{Sec:invariant}
	
	In standard $4D$ Minkowski spacetime, considering the $(-,+,+,+)$ signature, the line element is defined as:
	
	\begin{equation}
		ds^2 = -c^2 dt^2 + dx^2 + dy^2 + dz^2
	\end{equation}
	
	We calculate the extended invariant interval $ds^2$ using two independent methods: the fundamental metric tensor approach and the algebraic coordinate expansion. Both methods must converge to the same result.
	
	\subsubsection*{Method 1: The Algebraic Approach}\label{Sec:met1}
	
	We derive the extended invariant interval by substituting the complex temporal coordinate $\tau = t + i\kappa$ directly into the standard Minkowski line element.
	
	To achieve this, the geometry must strictly utilize a symmetric bilinear form. Applying it to the complex-time coordinate yields:
	
	\begin{equation}
		(dt+id\kappa)^2 = (dt+id\kappa)(dt+id\kappa) = dt^2 - d\kappa^2 + 2idtd\kappa
	\end{equation}
	
	Taking the differential $d\tau = dt + i d\kappa$, we square the temporal term:
	
	\begin{align}
		ds^2 &= -c^2(d\tau)^2 + dx^2 + dy^2 + dz^2 \nonumber \\
		&= -c^2(dt + i d\kappa)^2 + dx^2 + dy^2 + dz^2 \nonumber \\
		&= -c^2(dt^2 + 2i dt d\kappa - d\kappa^2) + dx^2 + dy^2 + dz^2 \nonumber \\
		&= (-c^2 dt^2 + c^2 d\kappa^2 + dx^2 + dy^2 + dz^2) - 2ic^2 dt d\kappa
		\label{method1}
	\end{align}
	
	Consequently, the complexified invariant interval separates into a real component, $\text{Re}(ds^2)$, and an imaginary part, 	$\text{Im}(ds^2)$. 
	
	The real component is:
	
	\begin{equation}
		\text{Re}(ds^2) = -c^2 dt^2 + c^2 d\kappa^2 + dx^2 + dy^2 + dz^2
		\label{real_ds}
	\end{equation}
	
	As previously stated in Sec. \ref{Sec:twotime}, the auxiliary parameter $\kappa$ must possess the exact same SI unit as thermodynamical time $t$ (seconds) to preserve dimensional homogeneity. The invariant interval requires every term to yield units of meters squared $(m^2)$. Given that the speed of light $c$ is measured in $(\frac{m}{s})$ and the standard temporal term $c^2dt^2$ resolves to squared length $(\frac{m}{s} \cdot s)^2 = m^2$, the auxiliary parameter $\kappa$ is required to be measured in seconds.
	
	The imaginary component is:
	
	\begin{equation}
		\text{Im}(ds^2) = -2c^2 dt d\kappa
		\label{imag_ds}
	\end{equation}
	
	Isolating the real part of the extended line element is geometrically necessary when evaluating macroscopic causal boundaries in Sec. \ref{Sec:lighthyperbola}. This approach has been applied in pseudo-complex gravity theories \cite{Weber:2025oyk}, where the constraint of evaluating only the real part of the metric component is tied to the fact that macroscopic particles traverse real paths.
	
	The imaginary non-vanishing part is essential for this framework as it acts as the coupling term between the standard thermodynamical time $t$ and the auxiliary parameter $\kappa$. This will ultimately lead to the imaginary off-diagonal elements in the extended $5 \cross 5$ metric tensor.
	
	\subsubsection*{Method 2: The Tensor Approach}\label{Sec:met2}
	
	We will expand the standard line element $ds^2 = g_{\mu\nu} dx^\mu dx^\nu$ over the coordinates: $x^0 = ct, x^1 = x, x^2 = y, x^3 = z, x^4 = c\kappa$ :
	
	\begin{equation*}
		ds^2 = g_{00}(dx^0)^2 + g_{11}(dx^1)^2 + g_{22}(dx^2)^2 + g_{33}(dx^3)^2 + g_{44}(dx^4)^2 + (g_{04}dx^0dx^4 + g_{40}dx^4dx^0)
	\end{equation*}
	
	\par From the uncoupled diagonal terms $g_{00} = -1$, $g_{11} = g_{22} = g_{33} = g_{44} = +1$, the line element becomes: 
	
	\begin{equation}
		ds^2 = (-c^2 dt^2 + dx^2 + dy^2 + dz^2 + c^2 d\kappa^2) + (g_{04}dx^0dx^4 + g_{40}dx^4dx^0)
		\label{method2}
	\end{equation}
	
	In order for Eq. \eqref{method2} to be equivalent to Eq. \eqref{method1}, it is straightforward that: 
	
	\begin{equation}
		-2i c^2 dt d\kappa = g_{04}dx^0dx^4 + g_{40}dx^4dx^0
	\end{equation}
	
	Taking into account the fact that the metric tensor is symmetric ($g_{\mu \nu} = g_{\nu \mu}$), we obtain $g_{04} = g_{40}$:
	
	\begin{equation}
		-2i c^2 dt d\kappa = 2 g_{04} c^2 dt d\kappa
	\end{equation}
	
	Consequently, the off-diagonal terms in the 5D metric tensor are:
	
	\begin{equation}
		g_{04} = g_{40} = -i
	\end{equation}
	
	As a result, the coupled extended $5 \cross 5$ metric tensor is:
	
	\begin{equation}
		g_{\mu\nu} = \begin{pmatrix} -1 & 0 & 0 & 0 & -i \\ 0 & 1 & 0 & 0 & 0\\ 0 & 0 & 1 & 0 & 0 \\ 0 & 0 & 0 & 1 & 0  \\ -i & 0 & 0 & 0 & 1 \end{pmatrix}
		\label{coupled}
	\end{equation}
	
	Consequently, the off-diagonal terms $(-i)$ secure the coupling between our two variables, confirming the co-existence property.
	
	If one were to incorrectly assume that the uncoupled metric tensor shown in Eq. \eqref{uncoupled} is the true form, this would have resulted into the vanishing of the imaginary cross-term: $\text{Im}(ds^2) = -2ic^2 dt d\kappa = 0$. This condition would violate the dynamic phase transition and the co-existence particularity.
	
	Finally, although the extended metric introduces a five-coordinate space $(ct,x,y,z,c\kappa)$, it is essential to establish that the framework is constrained to four degrees of freedom. As seen in Eq. \eqref{tau-eq}, the variables $t$ and $\kappa$ are not independent, but they are geometric projections linked through the transition phase angle $\theta$. 
	
	\subsection{Properties of the Complex-Time Parametrization: Degeneracy and Non-perpendicularity}\label{Sec:properties}
	
	Having mathematically proved the co-existence of standard time $t$ and auxiliary parameter $\kappa$ through the off-diagonal terms shown in Eq. \eqref{coupled}, we can now address the two remaining properties of this extended framework: non-perpendicularity and degeneracy.
	
	While the imaginary element $i$ acts as a geometric rotating operator indicating orthogonality, the presence of the off-diagonal $-i$ terms in the extended metric tensor dictates a dynamic non-perpendicularity. This guarantees that the $t$ and $\kappa$ variables are not treated independently; furthermore, this non-perpendicularity between the temporal and auxiliary axes represents the non-Hermitian coupling between classical propagation and the evanescent state. In spite of that, this coupling alters the invertibility of the metric.
	
	Based on the form of the extended metric tensor, it is straightforward to evaluate its determinant:
	
	\begin{equation}
		\det(g_{\mu\nu}) = 
		\begin{vmatrix} 
			-1 & 0 & 0 & 0 & -i \\ 
			0 & 1 & 0 & 0 & 0 \\ 
			0 & 0 & 1 & 0 & 0 \\ 
			0 & 0 & 0 & 1 & 0 \\ 
			-i & 0 & 0 & 0 & 1 
		\end{vmatrix} 
		= \begin{vmatrix} 
			-1 & -i \\ 
			-i & 1 
		\end{vmatrix}
		= (-1)(1) - (-i)(-i) = -1 - (-1) = 0
	\end{equation}
	
	This demonstrates the ultimate particularity of the extended framework: degeneracy.
	
	Firstly, this shows that the imaginary auxiliary $i\kappa$ axis does not introduce extra degrees of freedom. Moreover, the degeneracy avoids ghost modes by mathematically erasing the independent kinetic term associated with the auxiliary $\kappa$ parameter.
	
	Secondly, one of the issues arising from this strict degeneracy is how we will define the inverse metric. Since the extended metric with determinant zero does not define an ordinary pseudo-Riemannian spacetime, the complex metric lacks a standard Lorentzian signature. 
	
	Furthermore, substituting the vanishing determinant into the inverse metric, which is defined as:
	
	\begin{equation}
		g^{\mu\nu} = \frac{1}{\det(g)} \cdot \text{adj}(g_{\mu\nu})
	\end{equation}
	
	would result into  $g^{\mu \nu} \propto \frac{1}{0}$, causing the standard inverse metric to diverge to infinity.
	
	Therefore, the standard operations used in Relativity, such as raising and lowering indices, constructing curvature tensors, defining a Levi-Civita connection, are not available without an additional geometric structure.
	
	We acknowledge that the extended complex-time metric is strictly degenerate. While this manuscript treats the $\kappa$ parameter as an auxiliary variable that phenomenologically describes spatial evanescence, the central question is whether a fully covariant general-relativistic extension is mathematically possible.
	
	As demonstrated by Garnier and Battista \cite{Garnier:2025jnp}, a covariant pseudoinverse can be constructed by introducing a fixed Riemannian metric $\zeta_{ab}$ and a set of covariant Moore-Penrose identities, which allows for the proper definition of curvature-related tensors. The full derivation of the extension from flat-space to curved-space is beyond the scope of this phenomenological paper and will be discussed in future work. 
	
	However, within the current flat-space parametrization, we can deduce the physical implication of this degeneracy by applying a Singular Value Decomposition (SVD) to the isolated $2 \cross 2$ complex-time part of the metric:
	
	\begin{equation}
		g_{complex-time} = \begin{pmatrix} -1 & -i \\ -i & 1 \end{pmatrix}
	\end{equation}
	
	 Solving the characteristic equation $\left(\det(g^*g - \sigma I) = 0 \right)$ yields the following singular values:
	
	\begin{equation}
		\sigma_1 = 2, \quad \sigma_2 = 0
		\label{SVD}
	\end{equation}
	
	The zero singular value ($\sigma_2 = 0$) demonstrates that the real thermodynamical time stops ($dt = 0$) during the transition. The remaining non-zero singular value ($\sigma_1 = 2$) guarantees that the surviving trajectory is projected entirely into the imaginary auxiliary axis ($i\kappa$).
	
	\subsection{5-Component Parametric Vectors}\label{Sec:vectors}
	
	Analogous to the 4-vectors in standard $4D$ Minkowski spacetime, we derive the formalism to 5-component parametric vectors within the extended complex-time parametrization framework. These 5-component entities strictly obey standard tensor algebraic rules.
	
	To construct the fundamental parametric vectors, we consider the coordinates: $x^0 = ct, x^1 = x, x^2 = y, x^3 = z, x^4 = c\kappa$
	
	\subsubsection{Extended Coordinate Vector}\label{Sec:coordinate}
	
	The primary entity is the contravariant extended coordinate vector, which specifies the particle's geometric location across spatial dimensions and complex temporal plane:
	
	\begin{equation}
		X^\mu = (ct, x, y, z, c\kappa)
	\end{equation}
	
	\subsubsection{Scalar Product}\label{Sec:scalarproduct}
	
	Before we define the effective complex-time velocity vector, we must establish how the scalar product will change accordingly to this extended complex-time framework.
	
	Taking into account the derivation of the metric tensor in Eq. \eqref{coupled}, the scalar product generally defined as:
	
	\begin{equation}
		A \cdot B = g_{\mu \nu} A^\mu B^\nu
	\end{equation}
	
	expands to:
	
	\begin{equation}
		A \cdot B = - A^0 B^0 + A^1 B^1 + A^2 B^2 + A^3 B^3 + A^4 B^4 + (-i) A^0 B^4 + (-i) A^4 B^0
	\end{equation}
	
	To analyze the geometric phase transition between real thermodynamical time and the auxiliary parameter, we evaluate the scalar product strictly within the complex temporal plane $(x^0,x^4)$. By setting the $3D$ spatial components to zero, we project the vectors into the spatial rest frame. This is justified because the extended signature flip during barrier penetration is a purely temporal rotation, independent of the macroscopic spatial axes.
	
	Hence, $(A \cdot B)_{3D} = A^1 B^1 + A^2 B^2 + A^3 B^3 = 0$. Therefore, the scalar product is:
	
	\begin{equation}
		A \cdot B = - A^0 B^0 + A^4 B^4 + (-i) A^0 B^4 + (-i) A^4 B^0
	\end{equation}
	
	This general formula will help us define the squared norms of the effective complex-time velocity $(U^2)$ and Momentum $(P^2)$.
	
	\subsubsection{Effective Complex-Time Velocity}\label{Sec:velocity}
	
	In standard $4D$ relativity, the 4-velocity vector of a massive particle is defined as the derivative of the 4-position vector with respect to the proper time $\tau_\text{proper}$. This is because $\tau_\text{proper}$ is a true Lorentz scalar and it has the same numerical value for all observers in any inertial reference frame:
	
	\begin{equation}
		U^\mu = \frac{d X^\mu}{d \tau_\text{proper}}
	\end{equation}
	
	Moreover, the invariant proper time interval is derived directly from the line element $(ds^2 = - c^2 d\tau_\text{proper}^2)$:
	
	\begin{equation}
		\Delta \tau_\text{proper}^2 = \Delta t^2 - \frac{\Delta x^2 + \Delta y^2 + \Delta z^2}{c^2}
		\label{proper_time_normal}
	\end{equation}
	
	However, as shown in Eq. \eqref{SVD}, our SVD analysis demonstrates that, during quantum tunneling traversal, the standard thermodynamical time pauses $(dt = 0)$. A direct consequence of this aspect is that proper time is also pausing $(d \tau_\text{proper} = 0)$. This implies that $\tau_\text{proper}$ cannot be a valid affine parameter for the extended complex-time framework.
	
	We must define a new invariant parameter to track the trajectory where standard proper time is zero. This affine parameter is defined as the complex-time magnitude $\left| \tau \right|$ which is composed of two Lorentz scalars: the standard invariant proper time $\tau_\text{proper}$ and the auxiliary parameter $\kappa$.
	
	As established in Sec. \ref{Sec:scalarproduct}, the geometric phase transition is evaluated in the particle's spatial rest frame $(dx = dy = dz = 0)$. This means that, based on Eq. \eqref{proper_time_normal}, the coordinate time $t$ is mathematically identical to proper time $(t = \tau_\text{proper})$. 
	
	By defining the auxiliary parameter $\kappa$ as a frame-independent property of the barrier penetration, we construct the invariant magnitude $\left| \tau \right|$, which will serve as the affine parameter within our extended framework. 
	
	\begin{equation}
		\left| \tau \right| = \sqrt{\tau_\text{proper}^2 + \kappa^2}
	\end{equation}
	
	Expanding Eq. \eqref{proper_time_normal} according to the five-coordinate framework, yields the macroscopic interval for the effective complex-time parameter:
	
	\begin{equation}
		d \left| \tau \right|^2 = d t^2 + d \kappa^2 - \frac{d x^2 + d y^2 + d z^2}{c^2}
		\label{extended_proper}
	\end{equation}
	
	Since we will study how this 5-Component Parametric Vectors behave when applied to the quantum tunneling process, it is easy to observe that in the classical regions outside the barrier, Eq. \eqref{extended_proper} reduces to the standard Lorentz proper time interval shown in Eq. \eqref{proper_time_normal} $(d\kappa = 0 \rightarrow d \left| \tau \right|^2 = d \tau_\text{proper}^2)$.
	
	Coming back to our extended vector derivation, the general formula for the 5-component effective velocity vector is:
	
	\begin{equation}
		U^\mu = \frac{d X^\mu}{d \left| \tau \right|}
	\end{equation}
	
	Considering the polar form in Eq. \eqref{polar_form}, we can parametrize the evolution in the rest frame as:
	
	\begin{equation}
		\tau_\text{proper} = \left| \tau \right| cos(\theta)
		\label{realtimeproper}
	\end{equation}
	
	and 
	
	\begin{equation}
		\kappa = \left| \tau \right| sin(\theta)
		\label{kappa}
	\end{equation}
	 
	Applying the product rule and $\frac{d \left| \tau \right|}{d \left| \tau \right|} = 1$, we take the derivative with respect to $|\tau|$ for both real and imaginary parametric terms:
	
	\begin{equation}
		\frac{d\tau_\text{proper}}{d \left| \tau \right|} =  cos(\theta) - \left| \tau \right| \cdot sin(\theta) \cdot \frac{d\theta}{d \left| \tau \right|}
	\end{equation}
	
	\begin{equation}
		\frac{d\kappa}{d \left| \tau \right|} =  sin(\theta) + \left| \tau \right| \cdot cos(\theta) \cdot \frac{d\theta}{d \left| \tau \right|}
	\end{equation}
	
	For simplicity, we will only calculate the $U^0$ and $U^4$ components of the effective complex-time velocity vector:
	
	\begin{equation}
		U^0 = \frac{dX^0}{d \left| \tau \right|} = \frac{d(ct)}{d \left| \tau \right|} = c \frac{dt}{d \left| \tau \right|} = c \cdot \left( cos(\theta) - \left| \tau \right| \cdot sin(\theta) \cdot \frac{d\theta}{d \left| \tau \right|} \right)
		\label{U_0}
	\end{equation}
	
	\begin{equation}
		U^4 = \frac{dX^4}{d \left| \tau \right|} = \frac{d(c\kappa)}{d \left| \tau \right|} = c \frac{d \kappa}{d \left| \tau \right|} = c \cdot \left( sin(\theta) + \left| \tau \right| \cdot cos(\theta) \cdot \frac{d\theta}{d \left| \tau \right|} \right)
		\label{U_4}
	\end{equation} 
	
	We evaluate $U^2$ using the scalar product:
	
	\begin{align}
		U^2 &= g_{\mu \nu} U^\mu U^\nu \nonumber \\
		&= g_{00} \cdot (U^0)^2 + g_{44} \cdot (U^4)^2 + g_{04} \cdot U^0 U^4 + g_{40} \cdot U^4 U^0 \nonumber \\
		&= - (U^0)^2 + 	(U^4)^2 - 2i U^0 U^4 \nonumber \\
		&= - [(U^0)^2 - (U^4)^2 + 2i U^0 U^4] \nonumber \\
		&= - (U^0 + i \cdot U^4)^2 
		\label{U_squared}
	\end{align}
	
	Using the formulas for $U^0$ and $U^4$ from Eqs. \eqref{U_0} and \eqref{U_4} and Euler's relations, it is easy to observe that:
	
	\begin{equation}
		U^0 + i \cdot U^4 = c \cdot e^{i \theta} \left( 1 + i \left| \tau \right| \frac{d\theta}{d \left| \tau \right|}\right)
	\end{equation}
	
	Substituting this into Eq. \eqref{U_squared}, yields:
	
	\begin{equation}
		U^2 = - c^2 \cdot e^{2i\theta} \left[ 1 - \left| \tau \right|^2 \cdot \left( \frac{d \theta}{d \left| \tau \right|} \right)^2 + 2i \cdot \left| \tau \right| \cdot \frac{d \theta}{d \left| \tau \right|} \right]
		\label{finalUsquared}
	\end{equation}	
	
	We will study how this effective complex-time velocity norm behaves when explicitly applied to the tunneling process in Sec. \ref{Sec:tunneling} . 
	
	\subsubsection{Effective Complex-Time Momentum}\label{Sec:5-momentum}
	
	In standard $4D$ Special Relativity, the 4-momentum unifies energy and 3-dimensional momentum into a single mathematical object. It is generally defined as:
	
	\begin{equation}
		P^\mu =  \begin{pmatrix} \frac{E}{c} \\ \textbf{p} \end{pmatrix} =  \begin{pmatrix} \frac{E}{c} \\ p_x \\ p_y \\ p_z \end{pmatrix}
	\end{equation}
	
	Where $E$ is the total relativistic energy of the particle and \textbf{p} is the classical 3-momentum vector.
	
	Alternatively, the energy-momentum vector can also be defined using the rest mass and 4-velocity:
	
	\begin{equation}
		P^\mu = m U^\mu 
		\label{5momentum}
	\end{equation}
	
	It is essential to state that the norm of the 4-momentum vector is Lorentz invariant:
	
	\begin{equation}
		P^\mu P_\mu = - \frac{E^2}{c^2} + \left| \textbf{p} \right| ^2 = - m^2 c^2
		\label{standardmomentum}
	\end{equation}
	
	Considering the general formula for the 4-momentum vector shown in Eq. \eqref{5momentum} and the definition of the extended complex-time velocity vector in Eq. \eqref{U_0} and \eqref{U_4}, the temporal components of the effective complex-time momentum vector are:
	
	\begin{equation}
		P^0 = m \cdot U^0 = mc \left( \cos\theta - |\tau| \cdot \sin\theta \cdot \frac{d\theta}{d|\tau|} \right) 
		\label{P_0}
	\end{equation}
	
	\begin{equation}	
		P^4 = m \cdot U^4 = mc \left( \sin\theta + |\tau| \cdot \cos\theta \cdot \frac{d\theta}{d|\tau|} \right)
	\end{equation}
	
	In order to evaluate the norm of this vector, we have to lower the indices ($P_\mu = g_{\mu\nu} P^\nu$). Because the metric tensor has off-diagonal terms ($g_{04} = g_{40} = -i$), the real and imaginary momentum components become:
	
	\begin{equation}
		P_0 = g_{00} P^0 + g_{04} P^4 = -P^0 - i P^4
	\end{equation}
	
	\begin{equation}
		P_4 = g_{40} P^0 + g_{44} P^4 = - i P^0 + P^4
	\end{equation}
	
	Unlike the standard 4D case, where $P_0$ depends entirely on the relativistic energy $E$, in our extended framework, it is directly coupled to the auxiliary component $P^4$.
	
	In the extended framework, the squared norm of the effective complex-time momentum vector becomes:
	
	\begin{equation}
		P^2 = P_\mu P^\mu = m^2 U^2
	\end{equation}
	
	Since we have already derived the squared norm of the effective complex-time velocity vector in Eq. \eqref{finalUsquared}, $P^2$ is defined as:
	
	\begin{equation}
		P^2 = - m^2 c^2 \cdot e^{2i\theta} \left[ 1 - \left| \tau \right|^2 \cdot \left( \frac{d \theta}{d \left| \tau \right|} \right)^2 + 2i \cdot \left| \tau \right| \cdot \frac{d \theta}{d \left| \tau \right|} \right]
		\label{Psquared}
	\end{equation}
	
	This extended momentum norm is the central result of the parametrization. In standard quantum mechanics, tunneling enforces the classical 4-momentum to become purely imaginary, leading to paradoxes. Eq. \eqref{Psquared} solves these anomalies by demonstrating that momentum does not become arbitrarily imaginary, but it transitions continuously into the complex plane. We will explicitly study how this effective complex-time momentum norm behaves when applied to the quantum tunneling process in Sec. \ref{Sec:tunneling} .

	\subsubsection{Generalized Complex-Time Lorentz Factor}\label{Sec:lorentz}
	
	The Lorentz factor is a fundamental dimensionless quantity related to time dilation, length contraction and relativistic momentum and energy.
	
	The mathematical expression for $\gamma$ is:
	
	\begin{equation}
		\gamma = \frac{1}{\sqrt{1-\frac{v^2}{c^2}}}
	\end{equation}
	
	In the extended framework, the particle traverses the quantum tunneling by using the phase transition angle $\theta$ introduce by the complexification of time; therefore, we derive a generalized complex-time Lorentz factor, denoted as $\gamma_5$ using two independent methods: one using relativistic energy and the other one using the proper time (in our case, the affine parameter $\left| \tau \right|$).
	
	\begin{enumerate}
		\item \textbf{Method 1: Energy}
		
		The standard Lorentz factor can be derived directly from the principles of relativistic energy using the Einstein Mass-Energy Equivalence.
		
		Let $m$ the rest mass of the particle and $E$ the total energy. The generalized formula is:
		
		\begin{equation}
			E = \gamma m c^2
		\end{equation}
		
		Isolating $\gamma$ from the previous equation, yields:
		
		\begin{equation}
			\gamma = \frac{E}{m c^2}
		\end{equation}
		
		In the extended complex-time framework, the relativistic energy equation becomes:
		
		\begin{equation}
			E_5 = c P^0  
		\end{equation}
		
		Using the formula for $P^0$ calculated in Eq. \eqref{P_0}, we obtain: 
		
		\begin{equation}
			E_5 = mc^2 \left( \cos\theta - |\tau| \cdot \sin\theta \cdot \frac{d\theta}{d|\tau|} \right)
		\end{equation}
		
		Therefore, the generalized Lorentz $\gamma_5$ factor is:
		
		\begin{equation}
			\gamma_5 = \cos\theta - |\tau| \cdot \sin\theta \cdot \frac{d\theta}{d|\tau|}
			\label{gammafive}
		\end{equation}
		
		\item \textbf{Method 2: Proper time}
		
		In standard $4D$ Special Relativity, the relationship between coordinate time $dt$ and proper time $d\tau_{\text{proper}}$ is:
		
		\begin{equation}
			\frac{dt}{d\tau_\text{proper}} = \frac{1}{\sqrt{1-\frac{v^2}{c^2}}}
		\end{equation}
		
		This ratio is exactly the Lorentz factor:
		
		\begin{equation}
			\gamma = \frac{dt}{d\tau_{\text{proper}}}
		\end{equation}
		
		In the extended complex-time framework, the true invariant scalar is not proper time $\tau_{\text{proper}}$, but the complex-time magnitude $\left| \tau \right|$. 
		
		Consequently, the generalized $\gamma_5$ Lorentz factor is defined as:
		
		\begin{equation}
			\gamma_5 = \frac{dt}{d|\tau|}
		\end{equation}
		
		Using Eq. \eqref{U_0}, we find:
		
		\begin{equation}
			U^0 = \frac{dX^0}{d|\tau|} 
			= \frac{d(ct)}{d|\tau|} = c \cdot \frac{dt}{d|\tau|} 
		\end{equation}
		
		Rearranging for the temporal ratio yields:
		
		\begin{equation}
			\frac{dt}{d|\tau|} = \frac{U^0}{c} 
		\end{equation}
		
		Therefore, the generalized Lorentz $\gamma_5$ factor is identical to the result derived via the energy method in Eq. \eqref{gammafive}:
		
		\begin{equation}
			\gamma_5 = \frac{dt}{d|\tau|} = \frac{U^0}{c} = \cos\theta - |\tau| \cdot \sin\theta \cdot \frac{d\theta}{d|\tau|}
		\end{equation}
		
	\end{enumerate}
	
	In conclusion, having established the extended 5-Component Parametric Vector formalism, we now possess the mathematical toolkit required to evaluate quantum phenomena. In Sec.  \ref{Sec:tunneling}, we will apply these generalized formulas to the physical boundaries of quantum tunneling, where the temporal phase rotates into the fourth quadrant, triggering the geometric signature flip that resolves the imaginary 4-momentum paradox.
	
	\subsection{The Classical Macroscopic Limit}\label{Sec:classiclimit}
	
	In order to satisfy the Correspondence Principle, the extended complex-time framework must reduce to standard $4D$ Minkowski kinematics under macroscopic conditions.
	
	As established in Sec. \ref{Sec:properties}, macroscopic entities are bound to thermodynamical time evolution. Therefore, their temporal phase angle is $\theta = 0$ and experiences no phase rotation $\frac{d\theta}{d|\tau|} = 0$.
	
	Applying these classical limits to the imaginary auxiliary parameter in Eq. \eqref{kappa} results in:
	
	\begin{equation}
		\kappa = |\tau|\cdot sin(0) = 0
	\end{equation}
	
	This confirms that the auxiliary parameter $\kappa$ is physically inaccessible to macroscopic systems.
	
	In addition, we evaluate the temporal components of the effective complex-time velocity shown in Eqs. \eqref{U_0} and \eqref{U_4}: 
	
	\begin{equation}
		U^0 = c \cdot (cos(0) - |\tau| \cdot sin(0) \cdot 0) = c
	\end{equation}
	
	\begin{equation}
		U^4 = c \cdot (sin(0) + |\tau| \cdot cos(0) \cdot 0) = 0
	\end{equation}
	
	This reduces the complex-time velocity to the standard $4D$ rest-frame 4-velocity, where the temporal component is equal to the speed of light and the imaginary coordinate vanishes.
	
	Furthermore, applying the classical macroscopic limit to the squared extended complex-time momentum norm in Eq. \eqref{Psquared} yields:
	
	\begin{equation}
		P^2 = -m^2c^2 \cdot e^0 \cdot [ 1-|\tau|^2 \cdot (0)^2 + 2i \cdot |\tau| \cdot 0 ] = -m^2c^2
	\end{equation}
	
	This result reproduces the standard Lorentz invariant scalar for 4-momentum.
	
	Ultimately, because the generalized complex-time Lorentz factor calculated in Eq. \eqref{gammafive} was derived from the effective complex-time velocity vector projected into the particle's rest frame $(dx = dy = dz = 0)$, applying the macroscopic limit $(\theta = 0)$ collapses the equation to:
	
	\begin{equation}
		\gamma_5 = cos(0) - |\tau| \cdot sin(0) \cdot 0 = 1
	\end{equation}
	
	This coincides with the classical rest-frame standard Lorentz factor $(\gamma = 1)$.
	
	In conclusion, the extended complex-time framework fully preserves the fundamental principles of standard $4D$ Special Relativity. In the classical macroscopic limit, this generalized formalism collapses back into standard $4D$ Minkowski kinematics.
	
	\section{Hyperbolic Sections of the Complexified Null Boundary}\label{Sec:hyperbolic}
	
	This section explores how the causal boundaries of the lightcone change in the extended complex-time parametrization framework. By substituting the complex-time coordinate $\tau = t + i\kappa$ into the classical Minkowski metric, we demonstrate the transformation of causality.
	
	\subsection{The Classical (1+1)D Causal Baseline}\label{Sec:standardlightcone}
	
	In standard $4D$ Special Relativity, causal relations are typically defined by the lightcone. 
	
	The lightcone defines the boundaries of causality, separating events that can affect a specific point from those that cannot. Since nothing can travel faster than light ($v \leq c$), spacetime is split into three distinct regions:
	
	\begin{enumerate}
		\item \textbf{Timelike region ($ds^2 < 0$):} Inside the cone indicating causally connected events.
		\item \textbf{Lightlike region ($ds^2 = 0$):} The exact boundary region, where light travels.
		\item \textbf{Spacelike region ($ds^2 > 0$):} Outside the cone, which defines disconnected, acausal events in the "Elsewhere" region. 
	\end{enumerate}
	
	Initially, we consider the $4D$ Minkowski line element:
	
	\begin{equation}
		ds^2 = -c^2 dt^2 + dx^2 + dy^2 + dz^2
	\end{equation}
	
	\begin{enumerate}
		\item \textbf{Timelike region} 
		
		This region represents the inside of the lightcone which contains the causal past and causal future. 
		
		\begin{equation}
			ds^2 < 0 \Longrightarrow -c^2dt^2 + dx^2 < 0 \iff dx^2 < c^2dt^2
		\end{equation}
		
		In the Timelike region, the physical distance is less than the distance light would travel in a given time interval. Thus, signals can propagate at $v<c$ to establish a causal link.  
		
		\item	\textbf{Lightlike Region} 
		
		The Null region is the boundary where light or photons ($v = c$) can travel. 
		
		\begin{equation}
			ds^2 = 0 \Longrightarrow -c^2dt^2 + dx^2 = 0 \implies dx^2 = c^2dt^2
		\end{equation}
		
		\item \textbf{Spacelike Region} 
		
		Events in the Elsewhere region are causally disconnected, because the spatial distance exceeds the distance light could travel in a given time interval. .
		
		\begin{equation}
			ds^2 > 0 \Longrightarrow -c^2dt^2 + dx^2 > 0 \implies dx^2 > c^2dt^2
		\end{equation}
		
	\end{enumerate}
	
	The geometric implications of these equations are best understood by mapping them directly onto the causal structure of the lightcone:
	
	\begin{figure}[htbp]
		\centering
		\begin{tikzpicture}
			
			\draw[->] (-3.5, 0) -- (3.5, 0) node[right] {$x$};
			\draw[->] (0, -3) -- (0, 3) node[above] {$ct$};
			
			\draw[thick] (-2.5, -2.5) -- (2.5, 2.5);
			\draw[thick] (-2.5, 2.5) -- (2.5, -2.5);
			
			\draw (0, 2) ellipse (2cm and 0.3cm);
			\node at (0.8, 2.7) {future};
			
			\node at (2.7, 2.7) {$ds^2=0$};
			
			\draw (0, -2) ellipse (2cm and 0.3cm);
			\node at (0.8, -2.5) {past};
			
			\node at (0, -1.2) {$ds^2 < 0$};
			
			\fill (0,0) circle (2pt);
			\draw[<-] (0.2, 0.1) -- (1, 0.5) node[right] {origin/present};
			
			\node at (-2.3, 0.6) {elsewhere};
			\node at (2.5, -0.6) {elsewhere};
			\node at (2.5, -1.1) {$ds^2 > 0$};
			
		\end{tikzpicture}
		\caption{The classical lightcone in (1+1)D spacetime, illustrating the causal boundaries of Minkowski geometry. The origin separates the timelike regions from the Elsewhere region.}
	\end{figure}
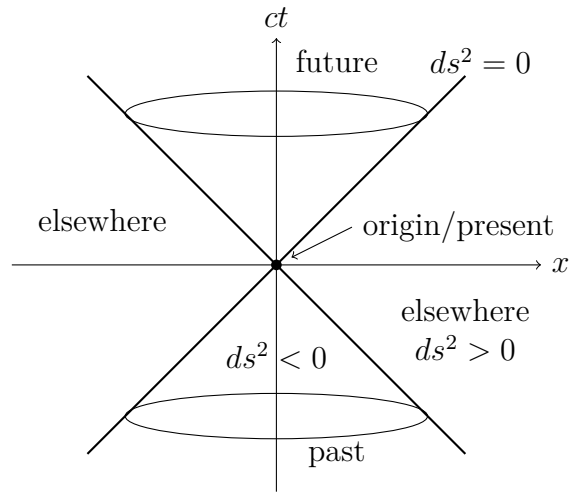
	
	\subsection{The Hyperbolic Section in the Complex Temporal Plane}\label{Sec:lighthyperbola}
	
	By expanding $4D$ Minkowski spacetime into a complex-time parametrization framework, the definitions of causal intervals must be reformulated. As derived in Eq. \eqref{method1}, the line element consists of both real and imaginary terms.
	
	While the non-vanishing imaginary term ($\text{Im}(ds^2) = -2c^2 dt d\kappa$) is necessary for this extended framework, it cannot be evaluated using the standard causal inequalities. Because the complex field is unordered, expressions such as $ds^2 < 0$ are undefined when $\text{Im}(ds^2) \neq 0$.
	
	Consequently, we adapt the classical causal boundaries to evaluate only the real component of the interval $(\text{Re}(ds^2))$ . By applying these causal conditions, we obtain a hyperbolic section of the higher-dimensional null cone.
	
	We consider the real part of the extended invariant interval:
	
	\begin{equation}
		\text{Re}(ds^2) = -c^2 dt^2 + c^2 d\kappa^2 + dx^2 + dy^2 + dz^2
	\end{equation}
	
	By suppressing the $y$ and $z$ spatial dimensions to evaluate a $2D$ spacetime cross-section, the causal boundaries are redefined as follows:
	
	\begin{enumerate}
		\item \textbf{Timelike Region}
		
		\begin{equation}
			\text{Re}(ds^2) < 0 \iff -c^2dt^2 + c^2d\kappa^2 + dx^2 < 0 \iff c^2d\kappa^2 + dx^2 < c^2dt^2
		\end{equation}
		
		Causality is still preserved, because the chronological evolution ($c^2 dt^2$) exceeds the system's combined spatial evolution ($dx^2$) and the evanescent depth ($c^2d\kappa^2$). The system remains anchored to classical causal reality.
		
		\item \textbf{Lightlike Region}
		
		\begin{equation} 
			\text{Re}(ds^2) = 0 \iff c^2dt^2 = c^2d\kappa^2 + dx^2
			\label{lightlike}
		\end{equation}
		
		In standard $4D$ spacetime, this region dictates the boundary defined by the speed limit. However, in our extended framework, at this exact boundary, a particle's temporal evolution is perfectly balanced by the combination of its physical spatial distance and auxiliary parameter $\kappa$. As we will later prove, the total extended complex-time velocity remains strictly bounded by the speed of light. 
		
		\item \textbf{Spacelike Region}
		
		\begin{equation}
			\text{Re}(ds^2) > 0 \iff -c^2dt^2 + c^2d\kappa^2 + dx^2 > 0 \iff c^2d\kappa^2 + dx^2 > c^2dt^2
		\end{equation}
		
		In standard relativity, a positive interval reveals a superluminal violation. However, in the extended framework, ($\text{Re} (ds^2) > 0$) mathematically describes a coordinate state where the system's traversal through the auxiliary parameter ($d\kappa$) and physical distance ($dx$) eclipses the real thermodynamical timeline. The trajectory has shifted orthogonally away from standard time, progressively aligning itself with the imaginary auxiliary axis.
		
	\end{enumerate}
	
	As previously stated, a consequence of the extended framework is the transformation of the lightcone into a light hyperbola.
	
	By integrating the null threshold boundary Eq. \eqref{lightlike} into macroscopic coordinates and evaluating it at a constant, non-zero auxiliary depth ($\kappa = \kappa_0$), we obtain:
	
	\begin{equation}
		c^2 t^2 - x^2 = c^2 \kappa_0 ^2
		\label{null}
	\end{equation}
	
	The mathematical definition of a vertical hyperbola is:
	
	\begin{equation}
		\frac{y^2}{a^2}-\frac{x^2}{b^2}=1
	\end{equation}
	
	By dividing Eq. \eqref{null} by $c^2 \kappa_0 ^2$, we obtain the exact complexified hyperbola equation:
	
	\begin{equation}
		\frac{t^2}{k_0^2} - \frac{x^2}{(ck_0)^2} = 1
	\end{equation}
	
	It is essential to observe that in a two-dimensional spacetime cross-section $(t, x)$, the classical causal boundary - previously defined by two intersecting null straight lines - has transformed into a light hyperbola. When expanded into three-dimensional coordinate space $(t, x, y)$, this causal surface formally manifests as a hyperboloid of two sheets.
	
	Because the hyperbola opens along the $t$-axis, the upper sheet represents the Absolute Future and the lower sheet the Absolute Past. The gap between them is not a void, but the suspension of thermodynamical time. It represents the duration where the real thermodynamic time has paused because the system has dived into the orthogonal auxiliary axis ($\kappa_0$). 
	
	The light hyperbola has a dynamic geometry dictated by the two fundamental parameters: $\kappa_0$ and $c$. 
	
	The parameter $\kappa_0$ represents the maximum evanescent depth achieved by the system. Because it is neither a universal constant nor a fixed value, it changes continuously. From a geometric perspective, $\kappa_0$ represents the position of the hyperbola's vertex along the real time axis. However, it is important to notice that no matter how large $\kappa_0$ becomes, the hyperbola remains strictly bounded by its two asymptotic cones: 
	
	\begin{equation}
		x \approx \pm ct \quad (\text{as } t, x \to \infty)
	\end{equation}
	
	The speed of light ($c$) dictates the fixed "openness" of these asymptotic boundaries. Therefore, even though the hyperbola's vertex may shift outward along the time axis, its trajectory will always curve asymptotically towards the limits defined by $c$. 
	
	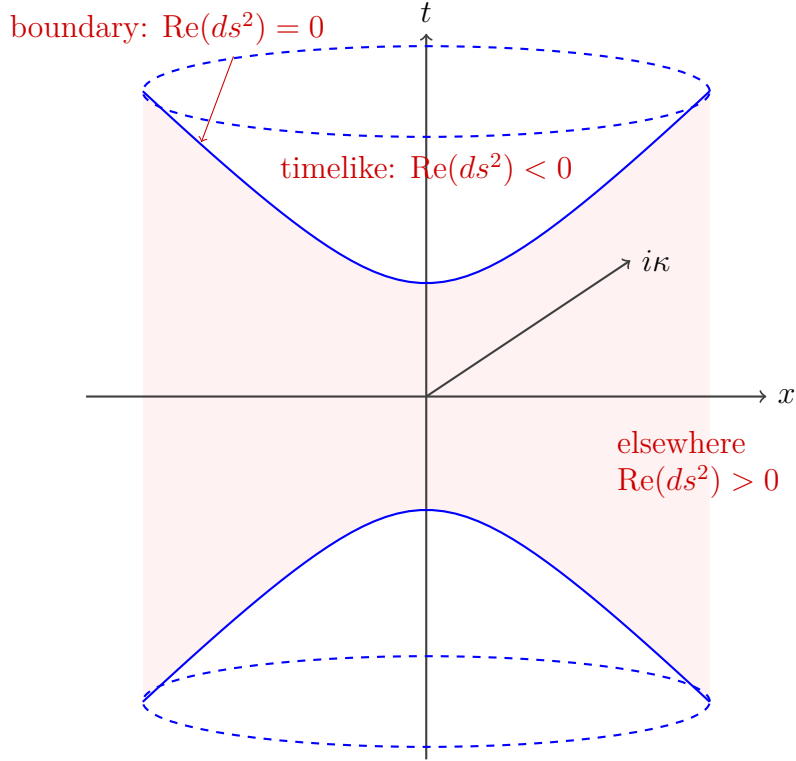
\begin{figure}[htbp]
		\centering
		\begin{tikzpicture}[scale=1.5]
			
			\fill[red!5] 
			plot[domain=-2.5:2.5, samples=50] ({\x}, {sqrt(1+\x*\x)}) -- 
			plot[domain=2.5:-2.5, samples=50] ({\x}, {-sqrt(1+\x*\x)}) -- cycle;
			
			\draw[->, thick, darkgray] (-3, 0) -- (3, 0) node[right, black] {$x$};
			\draw[->, thick, darkgray] (0, -3.2) -- (0, 3.2) node[above, black] {$t$};
			\draw[->, thick, darkgray] (0, 0) -- (1.8, 1.2) node[right, black] {$i\kappa$};
			
			\draw[thick, blue, domain=-2.5:2.5, samples=50] plot ({\x}, {sqrt(1+\x*\x)});
			\draw[thick, blue, domain=-2.5:2.5, samples=50] plot ({\x}, {-sqrt(1+\x*\x)});
			
			\draw[dashed, blue, thick] (0, 2.69) ellipse (2.5 and 0.4);
			\draw[dashed, blue, thick] (0, -2.69) ellipse (2.5 and 0.4);
			
			\node[text=red!80!black] at (0, 2) {timelike: $\text{Re}(ds^2) < 0$};
			
			\node[text=red!80!black, anchor=south east] at (-0.8, 3) {boundary: $\text{Re}(ds^2)=0$};
			\draw[->, red!80!black, shorten >=2pt] (-1.7, 3) -- (-2, 2.2);
			
			\node[text=red!80!black, align=left] at (2.4, -0.6) {elsewhere \\ $\text{Re}(ds^2) > 0$};
			
		\end{tikzpicture}
		\caption{Light Hyperbola}
		\label{hyperbola}
	\end{figure}
	
	Ultimately, satisfying the Correspondence Principle, the condition $\kappa_0 = 0$ dictates that the hyperbola's vertex returns to the origin, transforming it into the standard lightcone. Consequently, the expanded complex-time metric collapses into the classical $(3+1)$ Minkowski metric.
	
	The expansion of the lightcone into light hyperbola does not violate the speed of light, rather, it expands its domain. Using the null boundary condition shown in Eq. \eqref{null} and dividing by $dt^2$, we obtain the complex-time velocity limit:
	
	\begin{equation}
		c^2 = v^2 + c^2 \left( \frac{d\kappa}{dt}\right)^2
	\end{equation}
	
	In the extended framework, the speed of light $c$ changes into a combined spacetime velocity limit. This proves that the spatial velocity ($v = \frac{dx}{dt}$) must decrease when the particle traverses into the auxiliary domain. Moreover, $c$ remains the upper bound for the total velocity, preserving causality. 
	
	Based on Fig. \ref{hyperbola}, the major change is that the physical origin (the absolute present) is entirely swallowed by the acausal Elsewhere region.
	
	This condition is mathematically defined as:
	
	\begin{align}
		\text{Re}(s^2) > 0 &\iff -c^2t^2 + x^2 + c^2\kappa_0^2 > 0 \nonumber \\
		&\iff c^2t^2 - x^2 < c^2\kappa_0^2 \nonumber \\
		&\iff \frac{t^2}{\kappa_0^2} - \frac{x^2}{(c\kappa_0)^2} < 1
	\end{align}
	
	Evaluating this inequality at the exact origin of physical spacetime $(t=0, x=0)$, we find that $0 < 1$, which is always true. This geometrically confirms that a system with $(\kappa_0 \neq 0)$ is separated from its own classical starting point by a spacelike interval.
	
	In conclusion, rather than providing an absolute geometric proof of hidden locality, this framework suggests a speculative geometric interpretation of nonlocal-looking quantum behaviour.
	
	\section{Quantum Tunneling}\label{Sec:tunneling}
	
	The quantum tunneling effect is one of the most significant discoveries of 20-th century physics. In quantum mechanics, tunneling is the phenomena where a microscopic particle can travel a potential barrier, even when its total relativistic energy $(E)$ is lower than the potential barrier height $(V_0)$. In order for this traversal to be mathematically possible inside the barrier, the classical 4-momentum appears to become imaginary. Despite this, real physical measurement of a true Hermitian momentum operator must yield real numbers, not imaginary values. \cite{Turok:2013dfa}
	
	To demonstrate how the extended complex-time framework and the 5-Component Parametric Vectors reveal an alternative method to approach this process, we must first establish the standard quantum mechanics baseline.
	
	We consider an idealized, closed, isolated Hermitian system to guarantee that temporal phase evolution is strictly geometric, avoiding the external leakage associated with open, non-Hermitian environments.
	
	Let us consider a particle of mass $m$ and total energy $E$ approaching a rectangular potential barrier defined by:
	
	\begin{equation}
		V(x) = \begin{cases} 
			0 & \text{for } x < 0 \text{ (Region I: Outside the barrier)} \\
			V_0 & \text{for } 0 \leq x \leq a \text{ (Region II: Inside the barrier)} \\
			0 & \text{for } x > a \text{ (Region III: Outside the barrier)}
		\end{cases}
	\end{equation}
	
	This can be visualised geometrically:
	
	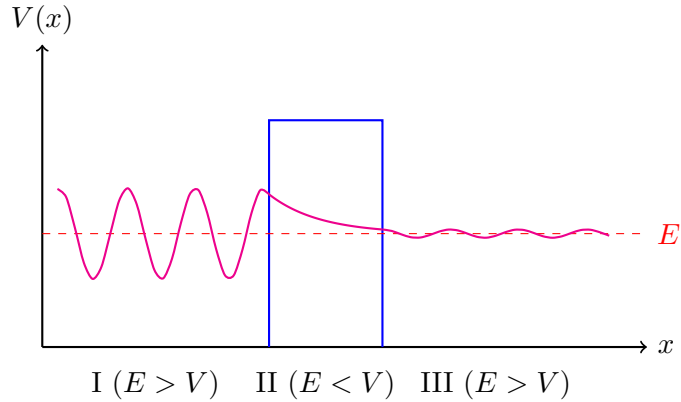
\begin{figure}[H]
		\centering
		\begin{tikzpicture}
			\draw[->, thick] (0,0) -- (8,0) node[right] {$x$};
			\draw[->, thick] (0,0) -- (0,4) node[above] {$V(x)$};
			
			\draw[thick, blue] (3,0) -- (3,3) -- (4.5,3) -- (4.5,0);
			
			\draw[dashed, red] (0,1.5) -- (8,1.5) node[right] {$E$};
			
			\draw[domain=0.2:3, smooth, variable=\x, magenta, thick] plot ({\x}, {1.5 + 0.6*sin(400*\x)});
			\node at (1.5, -0.5) {I ($E > V$)};
			
			\draw[domain=3:4.5, smooth, variable=\x, magenta, thick] plot ({\x}, {1.5 + 0.6*sin(400*3)*exp(-1.5*(\x-3))});
			\node at (3.75, -0.5) {II ($E < V$)};
			
			\draw[domain=4.5:7.5, smooth, variable=\x, magenta, thick] plot ({\x}, {1.5 + (0.6*sin(400*3)*exp(-1.5*1.5))*sin(400*(\x-4.5) + 90)});
			\node at (6, -0.5) {III ($E > V$)};
		\end{tikzpicture}
		\caption{Quantum Tunneling. The 3 distinct regions the particle traverses.}
	\end{figure}
	
	The particle's behavior is governed by the time-independent Schr\"{o}dinger's equation:
	
	\begin{equation}
		-\frac{\hbar^2}{2m} \frac{d^2\psi}{dx^2} + V(x)\psi = E\psi
	\end{equation}
	
	In standard $4D$, inside the barrier, the kinetic energy $T = E - V_0$ has a negative value. This violates classical mechanics, which demands that kinetic energy must be positive. Consequently, the classical momentum becomes imaginary and represents a causal paradox:
	
	\begin{equation}
		p = i \sqrt{2m(V_0 - E)}
	\end{equation}

	As a result, the oscillating wave $e^{ikx}$ turns into an exponentially decaying state $e^{-\kappa x}$ $(k = i\kappa)$ inside the barrier. 
	
	However, by sectioning the tunneling process and mapping the spatial decay using the auxiliary parameter $\kappa$, we propose an alternative kinematic approach.
	
	We introduce the auxiliary parameter $\kappa$:
	
	\begin{equation}
		\kappa(x) = - \frac{\chi}{\omega} \cdot x
	\end{equation}
	
	Here, $x$ represents the penetration depth inside the barrier (measured in meters), $\chi$ is the spatial decay constant of the evanescent wave (measured in $m^{-1}$) and $\omega$ is the angular frequency of the particle (measured in $s^{-1}$). It is mathematically straightforward to observe that $\kappa$ is strictly measured in seconds.
	
	It is important to note that the trajectory of the particle is dictated by the phase angle $\theta$, which rotates continuously through the fourth quadrant $(0 \geq \theta \geq \frac{-\pi}{2})$ to represent wavefunction decay without violating causality.
	
	To evaluate the kinematic state of the particle across these regions, we utilize the general formulas for the squared norm of the effective complex-time velocity and momentum norm derived in Eqs. \eqref{finalUsquared} and \eqref{Psquared}:
	
	\begin{equation}
		U^2 = - c^2 \cdot e^{2i\theta} \left[ 1 - \left| \tau \right|^2 \cdot \left( \frac{d \theta}{d \left| \tau \right|} \right)^2 + 2i \cdot \left| \tau \right| \cdot \frac{d \theta}{d \left| \tau \right|} \right]
	\end{equation}	
	
	\begin{equation}
		P^2 = - m^2 c^2 \cdot e^{2i\theta} \left[ 1 - \left| \tau \right|^2 \cdot \left( \frac{d \theta}{d \left| \tau \right|} \right)^2 + 2i \cdot \left| \tau \right| \cdot \frac{d \theta}{d \left| \tau \right|} \right]
	\end{equation}
	
	The five regions are defined by the continuous rotation of the temporal axis:
	
	\begin{enumerate}
		\item \textbf{Region I:} The classical limit before the barrier $(\theta = 0)$
		\item \textbf{Region II:} The entrance boundary penetration $(0 > \theta > \frac{-\pi}{2})$
		\item \textbf{Region III:} The pure quantum zone inside the barrier $(\theta = \frac{-\pi}{2})$
		\item \textbf{Region IV:} The exit boundary penetration $(\frac{-\pi}{2} < \theta < 0)$
		\item  \textbf{Region V:} The classical limit after the barrier $(\theta = 0)$
	\end{enumerate}
	
	In the following section, we evaluate the wavefunction dynamics across these five regions. 
	
	\begin{enumerate}
		\item \textbf{Region I: The classical limit before the barrier $(\theta = 0)$}
		
		Outside the barrier, the particle's total energy exceeds the potential $(E > V)$.
		Because the temporal phase angle is zero $(\theta = 0)$, evaluating Eqs. \eqref{tau-eq} and \eqref{polar_form} at this limit yields a purely real temporal axis $(\tau = t, \kappa = 0)$. In addition, the rate of phase change with respect to the effective complex-time magnitude is zero: $\frac{d \theta}{d \left| \tau \right|} = 0$.
		
		Under these conditions, the effective complex-time velocity norm (Eq. \eqref{finalUsquared}) is equal to:
		
		\begin{equation}
			U^2 = - c^2 \cdot e^{2i \cdot 0} \left[ 1 - \left| \tau \right|^2 \cdot (0)^2 + 2i \cdot \left| \tau \right| \cdot 0 \right] = -c^2 \cdot 1 = -c^2
		\end{equation}	
		
		Evaluating the effective complex-time momentum norm shown in Eq. \eqref{Psquared} under this classical limit results in:
		
		\begin{equation}
			P^2 = -m^2c^2 \cdot e^0 \cdot \left[ 1-|\tau|^2 \cdot (0)^2 + 2i \cdot |\tau| \cdot 0 \right] = -m^2c^2
		\end{equation}
		
		This confirms that, outside the barrier, the particle's effective velocity and squared momentum are strictly negative, therefore the particle possesses a standard 4D macroscopic timelike signature.
		
		Furthermore, the temporal coordinates are strictly real $(t > 0, \kappa = 0)$ in the fourth quadrant when $\theta = 0$.
		
		Utilizing the energy relation $E = \hbar \omega$, the incident wavefunction remains purely oscillatory in standard thermodynamical time:
		
		\begin{equation}
			\psi (x,\tau) = \psi (x,t) = (A e^{ikx} + B e^{-ikx}) e^{-i\omega t}
			\label{reg1}
		\end{equation}
		
		\item \textbf{Region II: The Entrance Boundary Penetration $(0 > \theta > -\frac{\pi}{2})$}
		
		The particle enters a hybrid complex-time transition state, due to the continuous rotation dictated by the phase angle $(0 > \theta > -\frac{\pi}{2})$. This wall penetration phase is extremely important because there has to be a continuous geometric rotation into the fourth quadrant, before it switches to purely auxiliary temporal evolution  inside the quantum barrier.
		
		Because the phase angle is rotating dynamically from $0$ to $-\frac{\pi}{2}$, the derivative of the phase with respect to the effective complex-time magnitude is negative: $\frac{d \theta}{d \left| \tau \right|} < 0$.
		
		Evaluating the effective complex-time velocity norm results in:
		
		\begin{equation}
			U^2 = - c^2 \cdot e^{2i\theta} \left[ 1 - \left| \tau \right|^2 \cdot \left( \frac{d \theta}{d \left| \tau \right|} \right)^2 + 2i \cdot \left| \tau \right| \cdot \frac{d \theta}{d \left| \tau \right|} \right]
		\end{equation}	
		
		Substituting the initial conditions into the effective complex-time momentum norm yields:
		
		\begin{equation}
			P^2 = -m^2c^2 \cdot e^{2i\theta} \cdot \left[ 1-|\tau|^2 \cdot (\frac{d \theta}{d \left| \tau \right|})^2 + 2i \cdot |\tau| \cdot \frac{d \theta}{d \left| \tau \right|} \right] 
		\end{equation}
		
		Thus, both $U^2$ and $P^2$ evaluate to a complex value, due to the negative, non-zero last term of the equation.
		
		During this active phase rotation, both real time and auxiliary parameter co-exist: $t > 0$ and $\kappa < 0$, therefore the hybrid wavefunction becomes:
		
		\begin{equation}
			\psi (x, \tau) = \psi (x, t, \kappa) = (A e^{ikx} + B e^{-ikx}) e^{-i\omega t} e^{\omega \kappa}
		\end{equation}
		
		Unlike the pure oscillatory wavefunction in Region I, there is an additional term $e^{\omega \kappa}$. Since the rotation is bounded to the fourth quadrant, where $\kappa < 0$, this exponential term proves that the wave amplitude starts to transition into the auxiliary evanescent domain.
		
		\item \textbf{Region III: The Pure Quantum zone inside the barrier ($\theta = -\frac{\pi}{2}$)}
		
		Once the particle is fully inside the barrier, its total energy is less than the potential energy $(E < V)$. Since the temporal rotation is complete and the phase angle is locked at $\theta = -\frac{\pi}{2}$, there is no active rotation: $\frac{d \theta}{d \left| \tau \right|} = 0$.
		
		Using these conditions, the complex-time velocity norm (Eq. \eqref{finalUsquared}) yields the essential geometric signature flip:
		
		\begin{equation}
			U^2 = - c^2 \cdot e^{2i \cdot \pi} \left[ 1 - \left| \tau \right|^2 \cdot (0)^2 + 2i \cdot \left| \tau \right| \cdot 0 \right] = -c^2 \cdot (-1) = +c^2
		\end{equation}	
		
		Evaluating the complex-time momentum norm (Eq. \eqref{Psquared}) results in:
		
		\begin{equation}
			P^2 = -m^2c^2 \cdot e^{-i\pi} \cdot \left[ 1-|\tau|^2 \cdot (0)^2 + 2i \cdot |\tau| \cdot 0  \right] = -m^2 c^2 \cdot (-1) = +m^2c^2
		\end{equation}
		
		Consequently, inside the barrier, the particle's complex-time velocity and momentum norm squared are strictly positive. The system has undergone a continuous sign flip from a timelike signature $(-c^2)$ to a spacelike signature $(+c^2)$. This geometric transformation resolves the anomaly of imaginary 4-momentum observed in the standard quantum mechanics case.
		
		Furthermore, because $\theta = - \frac{\pi}{2}$, the real thermodynamical time component vanishes $(t = 0)$, resulting in purely auxiliary temporal evolution $(\tau = i \kappa)$. 
		
		Evaluating the partial time derivative under this condition yields: 
		
		\begin{equation}
			\pdv{\psi}{\tau} = \pdv{\psi}{i \kappa} = \frac{1}{i} \pdv{\psi}{\kappa} = (-i) \pdv{\psi}{\kappa}
		\end{equation}
		
		Substituting this into the generalized Schr\"{o}dinger equation provides:
		
		\begin{equation}
			i \hbar \pdv{\psi}{\tau} = \hat{H} \psi \Longrightarrow i \cdot (-i) \hbar \pdv{\psi}{\kappa} = \hat{H} \psi \Longrightarrow \hbar \pdv{\psi}{\kappa} = \hat{H} \psi
		\end{equation}
		
		This important result shows that in the pure quantum region, the wavefunction no longer oscillates through real time; instead, it turns into a purely decaying evanescent wave: 
		
		\begin{equation}
			\psi (x, \tau) = \psi (x, \kappa) = (C e^{ikx} + D e^{-ikx}) e^{\omega \kappa}
		\end{equation}
		
		\item \textbf{Region IV: The Exit Boundary Penetration} ($-\frac{\pi}{2} < \theta < 0$)
		
		As the particle reaches the exit boundary of the potential barrier, the evanescent depth accumulates to a maximum finite value ($\kappa = \kappa_{\text{total}}$). The temporal axis must rotate continuously back toward the real axis to re-enter macroscopic reality. 
		
		Because the phase angle is actively rotating from $-\frac{\pi}{2}$ toward $0$, the derivative of the phase is strictly positive: $\frac{d\theta}{d|\tau|} > 0$. 
		
		Similar to Region II, substituting this dynamic rotation back into the complex-time velocity and momentum norm results in a complex value because the term $2i|\tau|\frac{d\theta}{d|\tau|}$ is non-zero and positive:
		
		\begin{equation}
			U^2 = - c^2 \cdot e^{2i\theta} \left[ 1 - \left| \tau \right|^2 \cdot \left( \frac{d \theta}{d \left| \tau \right|} \right)^2 + 2i \cdot \left| \tau \right| \cdot \frac{d \theta}{d \left| \tau \right|} \right]
		\end{equation}	
		
		and
		
		\begin{equation}
			P^2 = -m^2c^2 \cdot e^{2i\theta} \cdot \left[ 1-|\tau|^2 \cdot (\frac{d \theta}{d \left| \tau \right|})^2 + 2i \cdot |\tau| \cdot \frac{d \theta}{d \left| \tau \right|} \right] 
		\end{equation}
		
		This mathematically defines Region IV as the inverse continuous transition phase, bridging the gap from a spacelike back to a timelike signature.
		
		Correspondingly, the hybrid wavefunction becomes:
		
		\begin{equation}
			\psi(x,\tau) = \psi(x,t,\kappa) = F e^{ikx} e^{-i\omega t} e^{\omega\kappa_{\text{total}}}
		\end{equation}
		
		\item \textbf{Region V: The classical limit after the barrier} ($\theta = 0$)
		
		Having completely traversed the barrier, the particle re-enters the macroscopic classical region where its total energy again exceeds the localized potential $(E > V)$. The temporal phase angle locks back onto the real axis ($\theta = 0$), stopping any further phase rotation ($\frac{d\theta}{d|\tau|} = 0$).
		
		Evaluating the complex-time velocity and momentum norm under these limits once again yields a purely timelike signature:
		
		\begin{equation}
			U^2 = - c^2 \cdot e^{2i \cdot 0} \left[ 1 - \left| \tau \right|^2 \cdot (0)^2 + 2i \cdot \left| \tau \right| \cdot 0 \right] = -c^2 \cdot 1 = -c^2
		\end{equation}	
		
		and
		
		\begin{equation}
			P^2 = -m^2c^2 \cdot e^0 \left[ 1 - |\tau|^2 \cdot (0)^2 + 2i \cdot |\tau| \cdot 0 \right] = -m^2c^2
		\end{equation}
		
		Because $\theta = 0$, the auxiliary depth does not evolve anymore, locking the accumulated depth at $\kappa_{\text{total}}$. 
		
		The wavefunction resumes pure standard thermodynamic oscillation:
		
		\begin{equation}
			\psi(x,\tau) = \psi(x,t) = \bar{F} e^{ikx} e^{-i\omega t}
		\end{equation}
		
		However, the geometric traversal through the auxiliary evanescent depth has left a permanent macroscopic footprint. The final transmitted amplitude ($\bar{F}$) is strictly less than the initial incident amplitude ($A$) due to the accumulated decay factor $e^{\omega\kappa_{\text{total}}}$, perfectly replicating the transmission coefficient of standard quantum mechanics without violating relativistic causality.
		
		Ultimately, this demonstrates that the wavefunction's transition from an oscillatory phase to a purely decaying state is mathematically governed by the replacement of real standard time evolution with spatial attenuation along the auxiliary axis. 
		
	\end{enumerate}
	
	In conclusion, the extended complex-time framework provides a rigorous geometric solution to the paradox of imaginary momentum in quantum tunneling. We have mathematically proved that barrier penetration is neither discontinuous or instantaneous; rather, the traversal is dictated by a continuous geometric rotation of the temporal axis, linked through the dynamic phase angle $\theta$. The imaginary 4-momentum anomaly is resolved through the continuous signature flip of the effective complex-time momentum norm. 
	
	\section{Empirical Evidence: Complex Transmission Delay in Scattering}\label{Sec:empirical}
	
	Having established the theoretical kinematics of this extended complex-time parametrization, we now turn to examining how this auxiliary evanescent depth can be phenomenologically compared to measurable complex time delays in open, non-Hermitian microwave scattering experiments.
	
	While our theoretical derivation in Sec. \ref{Sec:tunneling} relied on an idealized, isolated Hermitian system to isolate the pure geometric kinematics of tunneling, the complex-time parametrization can be connected to non-Hermitian environments.
	
	Recent microwave scattering experiments conducted by Giovannelli and Anlage \cite{Giovannelli:2024yxl} study complex transmission time delay in non-unitary scattering systems. By transmitting microwave pulses through a complex 2-port graph, they observed that the transmission delay is  not purely real, but it also possesses an imaginary component.
	
	In scattering theory, their complex transmission time delay $(\tau_{T})$ is derived from the rank 2 scattering matrix and is defined as:
	
	\begin{equation}
		\tau_{T} = -i \pdv{\omega} \ln S_{21}(\omega)
	\end{equation}
	
	The transmission time delay can be separated into its real and imaginary components:
	
	\begin{equation}
		\tau_T = Re[\tau_T] + i Im[\tau_T]
	\end{equation}
	
	This provides a possible phenomenological analogy for the auxiliary component of our proposed complex-time parameter:
	
	\begin{equation}
		d\tau = dt + i d\kappa 
	\end{equation}
	
	In our case, the real temporal evolution $(dt)$ is phenomenologically correlated to the real transmission delay component $(Re[\tau_T])$, while the auxiliary parameter $(d\kappa)$ is represented by the measured imaginary transmission delay $(iIm[\tau_T])$. It is essential to note that this empirical data offers a phenomenological relation where an imaginary time coordinate serves as a rigorous bookkeeping representation of spatial evanescence.
	
\begin{figure}[htbp]
		\centering
		\includegraphics[width=0.8\textwidth, height=0.3\textheight]{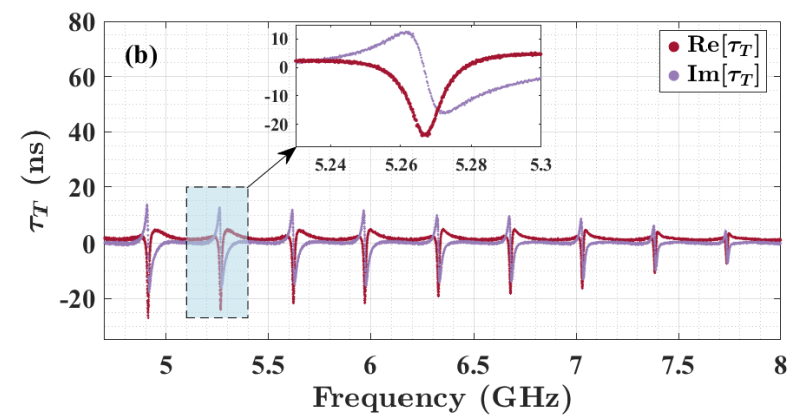} 
		\caption{Experimental measurement of complex transmission time delay in a microwave ring graph. Figure reproduced from Giovannelli and Anlage (2024).}
		\label{fig:Giov}
	\end{figure}

	The experimental data in Fig. \ref{fig:Giov} provides an observable analogue for the co-existence and dynamic rotation of these real and imaginary temporal components.
	
	We can directly map their observed resonance data to our distinct kinematic regions:
	
	\begin{enumerate}
		\item \textbf{The Classical Limit ($\theta = 0$)}
		
		In the off-resonance intervals (e.g., between 5.3 and 5.5 GHz), both $Re[\tau_T]$ and $Im[\tau_T]$ are approximately zero. In this null regime, the system does not undergo auxiliary spatial decay, but it propagates along the real thermodynamical time axis $(t > 0)$. The wave evolution is confined to standard classical mechanics and presents purely oscillatory behavior.
		
		\item \textbf{The Phase Transition ($0 > \theta > -\frac{\pi}{2}$)}
		
		In the proximity of the resonance peak (near 5.26 GHz), we observe the hybrid complex-time state. As the system begins to undergo the continuous temporal phase rotation, we see a simultaneous bifurcation - a rise and fall - in both temporal components, showing that the wave amplitude is transitioning into the auxiliary evanescent domain.
		
		\item \textbf{The Quantum Limit ($\theta = -\frac{\pi}{2}$)}
		
		At the exact point of resonance (5.26 GHz), the real time delay undergoes a sharp reversal along the negative axis, correlating to the theoretical state where real thermodynamical time is suspended ($dt = 0$). Simultaneously, the imaginary time delay presents a positive peak, confirming that the wave is fully described by the orthogonal, auxiliary spatial depth ($d\kappa > 0$). 
		
	\end{enumerate}
	
	Ultimately, while the proposed complex-time parametrization framework remains highly theoretical, these microwave scattering experiments demonstrate that our formalism can provide a physically observable analogy for analyzing evanescent decay and non-Hermitian wave scattering. Our prediction that introducing the complexification of time and a dynamic phase transition angle $\theta$ provides an alternative method to fully map wave dynamics inside the quantum barrier.
	
	\section{Conclusion}
	
	This research paper has introduced a complex-time parametrization proposed to explain the wave dynamics inside the quantum barrier traversal. We have demonstrated that formulating a complex-parametric vector formalism mathematically proves the geometric signature flip from a timelike to a spacelike state. By dividing the quantum penetration barrier into five distinct regions, we provided a mathematical demonstration that this traversal does not bypass classical relativistic rules nor the speed of light limit; instead, it can be modeled using a dynamic phase rotation angle that facilitates a continuous traversal.
	
	Furthermore, while this orthogonal auxiliary domain remains inaccessible from a standard macroscopic perspective due to real-time chronological evolution, our analysis proves that its traversal leaves an observable physical footprint. From a strictly phenomenological point, this theoretical framework is supported by empirical microwave scattering data, which correlates our proposed auxiliary coordinate with a measurable imaginary transmission time delay.
	
	Ultimately, the central focus of this framework revolves around the concept that the complexification of time can provide a consistent geometric resolution to the anomaly of imaginary momentum in quantum tunneling. 
	
	\section*{Author Contributions}
	The author confirms being the sole contributor of this work.
	
	\section*{Funding}
	This research received no external funding.
	
	\section*{Acknowledgement}
	
	I would like to express my sincere gratitude to my sister, Teodora Daia, for her constant support. I extend my thanks to my friends and colleagues Giorgio Foti and Iv\'{a}n Hern\'{a}ndez Garibay for their insightful exchanges. I am also deeply indebted to my colleagues within the FORQ group - Dr. Maxim Chernodub, Rajeev Singh, Sergio Morales-Tejera, Pracheta Singha, and Matteo Buzzegolli - for their constructive suggestions and fruitful discussions throughout the development of this manuscript. Finally, I am profoundly grateful to Prof. Alessandro Sergi for his critical reading of the early drafts and for providing the expert advice that significantly strengthened this work.
	
	\bibliography{bibliographyreferences}

\begin{thebibliography}{16}%
\makeatletter
\providecommand \@ifxundefined [1]{%
 \@ifx{#1\undefined}
}%
\providecommand \@ifnum [1]{%
 \ifnum #1\expandafter \@firstoftwo
 \else \expandafter \@secondoftwo
 \fi
}%
\providecommand \@ifx [1]{%
 \ifx #1\expandafter \@firstoftwo
 \else \expandafter \@secondoftwo
 \fi
}%
\providecommand \natexlab [1]{#1}%
\providecommand \enquote  [1]{``#1''}%
\providecommand \bibnamefont  [1]{#1}%
\providecommand \bibfnamefont [1]{#1}%
\providecommand \citenamefont [1]{#1}%
\providecommand \href@noop [0]{\@secondoftwo}%
\providecommand \href [0]{\begingroup \@sanitize@url \@href}%
\providecommand \@href[1]{\@@startlink{#1}\@@href}%
\providecommand \@@href[1]{\endgroup#1\@@endlink}%
\providecommand \@sanitize@url [0]{\catcode `\\12\catcode `\$12\catcode `\&12\catcode `\#12\catcode `\^12\catcode `\_12\catcode `\%12\relax}%
\providecommand \@@startlink[1]{}%
\providecommand \@@endlink[0]{}%
\providecommand \url  [0]{\begingroup\@sanitize@url \@url }%
\providecommand \@url [1]{\endgroup\@href {#1}{\urlprefix }}%
\providecommand \urlprefix  [0]{URL }%
\providecommand \Eprint [0]{\href }%
\providecommand \doibase [0]{https://doi.org/}%
\providecommand \selectlanguage [0]{\@gobble}%
\providecommand \bibinfo  [0]{\@secondoftwo}%
\providecommand \bibfield  [0]{\@secondoftwo}%
\providecommand \translation [1]{[#1]}%
\providecommand \BibitemOpen [0]{}%
\providecommand \bibitemStop [0]{}%
\providecommand \bibitemNoStop [0]{.\EOS\space}%
\providecommand \EOS [0]{\spacefactor3000\relax}%
\providecommand \BibitemShut  [1]{\csname bibitem#1\endcsname}%
\let\auto@bib@innerbib\@empty
\bibitem [{\citenamefont {Hund}(1927)}]{Hund1927}%
  \BibitemOpen
  \bibfield  {author} {\bibinfo {author} {\bibfnamefont {F.}~\bibnamefont {Hund}},\ }\bibfield  {title} {\bibinfo {title} {Zur deutung der molekelspektren. i},\ }\href {https://doi.org/10.1007/BF01400234} {\bibfield  {journal} {\bibinfo  {journal} {Zeitschrift fur Physik}\ }\textbf {\bibinfo {volume} {40}},\ \bibinfo {pages} {742} (\bibinfo {year} {1927})}\BibitemShut {NoStop}%
\bibitem [{\citenamefont {Steinberg}(1995)}]{Steinberg:1994ks}%
  \BibitemOpen
  \bibfield  {author} {\bibinfo {author} {\bibfnamefont {A.~M.}\ \bibnamefont {Steinberg}},\ }\bibfield  {title} {\bibinfo {title} {{How much time does a tunneling particle spend in the barrier region?}},\ }\href {https://doi.org/10.1103/PhysRevLett.74.2405} {\bibfield  {journal} {\bibinfo  {journal} {Phys. Rev. Lett.}\ }\textbf {\bibinfo {volume} {74}},\ \bibinfo {pages} {2405} (\bibinfo {year} {1995})},\ \Eprint {https://arxiv.org/abs/quant-ph/9501015} {arXiv:quant-ph/9501015} \BibitemShut {NoStop}%
\bibitem [{\citenamefont {Camus}\ \emph {et~al.}(2017)\citenamefont {Camus}, \citenamefont {Yakaboylu}, \citenamefont {Fechner}, \citenamefont {Klaiber}, \citenamefont {Laux}, \citenamefont {Mi}, \citenamefont {Hatsagortsyan}, \citenamefont {Pfeifer}, \citenamefont {Keitel},\ and\ \citenamefont {Moshammer}}]{Camus:2016yhc}%
  \BibitemOpen
  \bibfield  {author} {\bibinfo {author} {\bibfnamefont {N.}~\bibnamefont {Camus}}, \bibinfo {author} {\bibfnamefont {E.}~\bibnamefont {Yakaboylu}}, \bibinfo {author} {\bibfnamefont {L.}~\bibnamefont {Fechner}}, \bibinfo {author} {\bibfnamefont {M.}~\bibnamefont {Klaiber}}, \bibinfo {author} {\bibfnamefont {M.}~\bibnamefont {Laux}}, \bibinfo {author} {\bibfnamefont {Y.}~\bibnamefont {Mi}}, \bibinfo {author} {\bibfnamefont {K.~Z.}\ \bibnamefont {Hatsagortsyan}}, \bibinfo {author} {\bibfnamefont {T.}~\bibnamefont {Pfeifer}}, \bibinfo {author} {\bibfnamefont {C.~H.}\ \bibnamefont {Keitel}},\ and\ \bibinfo {author} {\bibfnamefont {R.}~\bibnamefont {Moshammer}},\ }\bibfield  {title} {\bibinfo {title} {{Experimental evidence for Wigner's tunneling time}},\ }\href {https://doi.org/10.1103/PhysRevLett.119.023201} {\bibfield  {journal} {\bibinfo  {journal} {Phys. Rev. Lett.}\ }\textbf {\bibinfo {volume} {119}},\ \bibinfo {pages} {023201} (\bibinfo {year} {2017})},\ \Eprint {https://arxiv.org/abs/1611.03701}
  {arXiv:1611.03701 [physics.atom-ph]} \BibitemShut {NoStop}%
\bibitem [{\citenamefont {Popov}(2005)}]{Popov:2005rp}%
  \BibitemOpen
  \bibfield  {author} {\bibinfo {author} {\bibfnamefont {V.~S.}\ \bibnamefont {Popov}},\ }\bibfield  {title} {\bibinfo {title} {{Imaginary-time method in quantum mechanics and field theory}},\ }\href {https://doi.org/10.1134/1.1903097} {\bibfield  {journal} {\bibinfo  {journal} {Phys. Atom. Nucl.}\ }\textbf {\bibinfo {volume} {68}},\ \bibinfo {pages} {686} (\bibinfo {year} {2005})}\BibitemShut {NoStop}%
\bibitem [{\citenamefont {Pisanty}\ and\ \citenamefont {Ivanov}(2014)}]{Pisanty:2014qre}%
  \BibitemOpen
  \bibfield  {author} {\bibinfo {author} {\bibfnamefont {E.}~\bibnamefont {Pisanty}}\ and\ \bibinfo {author} {\bibfnamefont {M.}~\bibnamefont {Ivanov}},\ }\bibfield  {title} {\bibinfo {title} {{Momentum transfers in correlation-assisted tunnelling}},\ }\href {https://doi.org/10.1103/PhysRevA.89.043416} {\bibfield  {journal} {\bibinfo  {journal} {Phys. Rev. A}\ }\textbf {\bibinfo {volume} {89}},\ \bibinfo {pages} {043416} (\bibinfo {year} {2014})},\ \Eprint {https://arxiv.org/abs/1309.4765} {arXiv:1309.4765 [physics.atom-ph]} \BibitemShut {NoStop}%
\bibitem [{\citenamefont {Klaiber}\ \emph {et~al.}(2023)\citenamefont {Klaiber}, \citenamefont {Can{\'a}rio},\ and\ \citenamefont {Hatsagortsyan}}]{Klaiber:2022ogb}%
  \BibitemOpen
  \bibfield  {author} {\bibinfo {author} {\bibfnamefont {M.}~\bibnamefont {Klaiber}}, \bibinfo {author} {\bibfnamefont {D.~B.}\ \bibnamefont {Can{\'a}rio}},\ and\ \bibinfo {author} {\bibfnamefont {K.~Z.}\ \bibnamefont {Hatsagortsyan}},\ }\bibfield  {title} {\bibinfo {title} {{Sub-barrier recollisions and the three classes of tunneling time delays in strong-field ionization}},\ }\href {https://doi.org/10.1103/PhysRevA.107.053103} {\bibfield  {journal} {\bibinfo  {journal} {Phys. Rev. A}\ }\textbf {\bibinfo {volume} {107}},\ \bibinfo {pages} {053103} (\bibinfo {year} {2023})},\ \Eprint {https://arxiv.org/abs/2208.10946} {arXiv:2208.10946 [physics.atom-ph]} \BibitemShut {NoStop}%
\bibitem [{\citenamefont {Schach}\ and\ \citenamefont {Giese}(2024)}]{Schach:2024kvq}%
  \BibitemOpen
  \bibfield  {author} {\bibinfo {author} {\bibfnamefont {P.}~\bibnamefont {Schach}}\ and\ \bibinfo {author} {\bibfnamefont {E.}~\bibnamefont {Giese}},\ }\bibfield  {title} {\bibinfo {title} {{A unified theory of tunneling times promoted by Ramsey clocks}},\ }\href {https://doi.org/10.1126/sciadv.adl6078} {\bibfield  {journal} {\bibinfo  {journal} {Sci. Adv.}\ }\textbf {\bibinfo {volume} {10}},\ \bibinfo {pages} {adl6078} (\bibinfo {year} {2024})},\ \Eprint {https://arxiv.org/abs/2404.14382} {arXiv:2404.14382 [quant-ph]} \BibitemShut {NoStop}%
\bibitem [{\citenamefont {Demir}\ and\ \citenamefont {Guner}(2017)}]{Demir:2017qkj}%
  \BibitemOpen
  \bibfield  {author} {\bibinfo {author} {\bibfnamefont {D.}~\bibnamefont {Demir}}\ and\ \bibinfo {author} {\bibfnamefont {T.}~\bibnamefont {Guner}},\ }\bibfield  {title} {\bibinfo {title} {{Statistical Approach to Tunneling Time in Attosecond Experiments}},\ }\href {https://doi.org/10.1016/j.aop.2017.09.009} {\bibfield  {journal} {\bibinfo  {journal} {Annals Phys.}\ }\textbf {\bibinfo {volume} {386}},\ \bibinfo {pages} {291} (\bibinfo {year} {2017})},\ \Eprint {https://arxiv.org/abs/1512.04338} {arXiv:1512.04338 [quant-ph]} \BibitemShut {NoStop}%
\bibitem [{\citenamefont {Peskin}\ and\ \citenamefont {Schroeder}(1995)}]{Peskin:1995ev}%
  \BibitemOpen
  \bibfield  {author} {\bibinfo {author} {\bibfnamefont {M.~E.}\ \bibnamefont {Peskin}}\ and\ \bibinfo {author} {\bibfnamefont {D.~V.}\ \bibnamefont {Schroeder}},\ }\href {https://doi.org/10.1201/9780429503559} {\emph {\bibinfo {title} {{An Introduction to quantum field theory}}}}\ (\bibinfo  {publisher} {Addison-Wesley},\ \bibinfo {address} {Reading, USA},\ \bibinfo {year} {1995})\BibitemShut {NoStop}%
\bibitem [{\citenamefont {Hess}\ and\ \citenamefont {Greiner}(2009)}]{Hess:2008wd}%
  \BibitemOpen
  \bibfield  {author} {\bibinfo {author} {\bibfnamefont {P.~O.}\ \bibnamefont {Hess}}\ and\ \bibinfo {author} {\bibfnamefont {W.}~\bibnamefont {Greiner}},\ }\bibfield  {title} {\bibinfo {title} {{Pseudo-complex General Relativity}},\ }\href {https://doi.org/10.1142/S0218301309012045} {\bibfield  {journal} {\bibinfo  {journal} {Int. J. Mod. Phys. E}\ }\textbf {\bibinfo {volume} {18}},\ \bibinfo {pages} {51} (\bibinfo {year} {2009})},\ \Eprint {https://arxiv.org/abs/0812.1738} {arXiv:0812.1738 [gr-qc]} \BibitemShut {NoStop}%
\bibitem [{\citenamefont {Carroll}(1997)}]{Carroll:1997ar}%
  \BibitemOpen
  \bibfield  {author} {\bibinfo {author} {\bibfnamefont {S.~M.}\ \bibnamefont {Carroll}},\ }\bibfield  {title} {\bibinfo {title} {{Lecture notes on general relativity}},\ }\href@noop {} {\  (\bibinfo {year} {1997})},\ \Eprint {https://arxiv.org/abs/gr-qc/9712019} {arxiv:gr-qc/9712019} \BibitemShut {NoStop}%
\bibitem [{\citenamefont {Misner}\ \emph {et~al.}(1973)\citenamefont {Misner}, \citenamefont {Thorne},\ and\ \citenamefont {Wheeler}}]{Misner:1973prb}%
  \BibitemOpen
  \bibfield  {author} {\bibinfo {author} {\bibfnamefont {C.~W.}\ \bibnamefont {Misner}}, \bibinfo {author} {\bibfnamefont {K.~S.}\ \bibnamefont {Thorne}},\ and\ \bibinfo {author} {\bibfnamefont {J.~A.}\ \bibnamefont {Wheeler}},\ }\href@noop {} {\emph {\bibinfo {title} {{Gravitation}}}}\ (\bibinfo  {publisher} {W. H. Freeman},\ \bibinfo {address} {San Francisco},\ \bibinfo {year} {1973})\BibitemShut {NoStop}%
\bibitem [{\citenamefont {Weber}\ \emph {et~al.}(2025)\citenamefont {Weber}, \citenamefont {Hess},\ and\ \citenamefont {Vasconcellos}}]{Weber:2025oyk}%
  \BibitemOpen
  \bibfield  {author} {\bibinfo {author} {\bibfnamefont {F.}~\bibnamefont {Weber}}, \bibinfo {author} {\bibfnamefont {F.~W. P.~O.}\ \bibnamefont {Hess}},\ and\ \bibinfo {author} {\bibfnamefont {C.~A.~Z.}\ \bibnamefont {Vasconcellos}},\ }\bibfield  {title} {\bibinfo {title} {{pc-Gravity as a geometric resolution of the black hole information paradox}},\ }\href {https://doi.org/10.1088/1402-4896/adfd9b} {\bibfield  {journal} {\bibinfo  {journal} {Phys. Scripta}\ }\textbf {\bibinfo {volume} {100}},\ \bibinfo {pages} {095301} (\bibinfo {year} {2025})},\ \Eprint {https://arxiv.org/abs/2506.21761} {arXiv:2506.21761 [gr-qc]} \BibitemShut {NoStop}%
\bibitem [{\citenamefont {Garnier}\ and\ \citenamefont {Battista}(2025)}]{Garnier:2025jnp}%
  \BibitemOpen
  \bibfield  {author} {\bibinfo {author} {\bibfnamefont {A.}~\bibnamefont {Garnier}}\ and\ \bibinfo {author} {\bibfnamefont {E.}~\bibnamefont {Battista}},\ }\bibfield  {title} {\bibinfo {title} {{Complex degenerate metrics in general relativity: a covariant extension of the Moore{\textendash}Penrose algorithm}},\ }\href {https://doi.org/10.1140/epjc/s10052-025-13957-w} {\bibfield  {journal} {\bibinfo  {journal} {Eur. Phys. J. C}\ }\textbf {\bibinfo {volume} {85}},\ \bibinfo {pages} {284} (\bibinfo {year} {2025})},\ \Eprint {https://arxiv.org/abs/2502.10053} {arXiv:2502.10053 [gr-qc]} \BibitemShut {NoStop}%
\bibitem [{\citenamefont {Turok}(2014)}]{Turok:2013dfa}%
  \BibitemOpen
  \bibfield  {author} {\bibinfo {author} {\bibfnamefont {N.}~\bibnamefont {Turok}},\ }\bibfield  {title} {\bibinfo {title} {{On Quantum Tunneling in Real Time}},\ }\href {https://doi.org/10.1088/1367-2630/16/6/063006} {\bibfield  {journal} {\bibinfo  {journal} {New J. Phys.}\ }\textbf {\bibinfo {volume} {16}},\ \bibinfo {pages} {063006} (\bibinfo {year} {2014})},\ \Eprint {https://arxiv.org/abs/1312.1772} {arXiv:1312.1772 [quant-ph]} \BibitemShut {NoStop}%
\bibitem [{\citenamefont {Giovannelli}\ and\ \citenamefont {Anlage}(2025)}]{Giovannelli:2024yxl}%
  \BibitemOpen
  \bibfield  {author} {\bibinfo {author} {\bibfnamefont {I.~L.}\ \bibnamefont {Giovannelli}}\ and\ \bibinfo {author} {\bibfnamefont {S.~M.}\ \bibnamefont {Anlage}},\ }\bibfield  {title} {\bibinfo {title} {{Physical Interpretation of Imaginary Time Delay}},\ }\href {https://doi.org/10.1103/nnk7-xy4v} {\bibfield  {journal} {\bibinfo  {journal} {Phys. Rev. Lett.}\ }\textbf {\bibinfo {volume} {135}},\ \bibinfo {pages} {043801} (\bibinfo {year} {2025})},\ \Eprint {https://arxiv.org/abs/2412.13139} {arXiv:2412.13139 [physics.optics]} \BibitemShut {NoStop}%
\end{thebibliography}%

\end{document}